\documentclass[reprint,aps,pre,groupedaddress,nobibnotes]{revtex4-1}
\usepackage{ microtype,enumitem,amsthm,amssymb,amsfonts,amsmath, amsbsy,amssymb,float,nicefrac}
\usepackage[stable]{footmisc}
\usepackage{listings}
\usepackage{graphicx}
\usepackage{caption}
\usepackage{subcaption}

\usepackage{appendix}
\usepackage[colorlinks=true]{hyperref}
\usepackage{mathtools,stackengine}

\usepackage{pgf,nicefrac}

\usepackage[textsize=scriptsize]{todonotes}

\usepackage[capitalize]{cleveref}

\begin{document}%
\title{The gift of gab: probing the limits of dynamic concentration-sensing across a network of communicating cells}
\author{Mohammadreza Bahadorian$^{1,2}$}
\author{Christoph Zechner$^{1,2,3}$}
\author{Carl Modes$^{1,2,3}$}
\affiliation{$1$ Max Planck Institut for Molecular Cell Biology and Genetics (MPI-CBG), 01037 Dresden, Germany}
\affiliation{$2$ Center for Systems Biology Dresden (CSBD), 01037 Dresden, Germany}
\affiliation{$3$ Cluster of Excellence Physics of Life, TU Dresden, 01307 Dresden, Germany.}
\begin{abstract}
Many systems in biology and beyond employ collaborative, collective communication strategies for improved efficiency and adaptive benefit. One such paradigm of particular interest is the community estimation of a dynamic signal, when, for example, an epithelial tissue of cells must decide whether to react to a given dynamic external concentration of stress signaling molecules. At the level of dynamic cellular communication, however, it remains unknown what effect, if any, arises from communication beyond the mean field level. What are the limits and benefits to communication across a network of neighbor interactions? What is the role of Poissonian vs. super Poissonian dynamics in such a setting? How does the particular topology of connections impact the collective estimation and that of the individual participating cells? In this letter we construct a robust and general framework of signal estimation over continuous time Markov chains in order to address and answer these questions. Our results show that in the case of Possonian estimators, the communication solely enhances convergence speed of the Mean Squared Error (MSE) of the estimators to their steady-state values while leaving these values unchanged. However, in the super-Poissonian regime, MSE of estimators significantly decreases by increasing the number of neighbors. Surprisingly, in this case, the clustering coefficient of an estimator does not enhance its MSE while reducing total MSE of the population. 
\end{abstract}

\maketitle

For cell populations and organisms, information about the surrounding environment can be life or death. Cells need to make good decisions in order to differentiate during development, protect themselves against fluctuating stresses, and make optimal use of their limited resources. An essential step of any cellular decision-making process is the inference of environmental signals, which are frequently encoded in the concentration of a stress molecule or ligand. Much effort has been directed to determine the properties and limits of concentration sensing by a single cell in steady-state \cite{berg1977physics,mora2015physical} as well as in dynamic conditions \cite{zechner2016molecular, kobayashi2010implementation}. 

Inference of environmental signals can be made by cells individually or collectively via cell-to-cell communication. In the latter case, the topology of interactions may play an important role in coupling and modulating the dynamics of individual cells.
The distributed estimation across communities has been a central focus of other fields \cite{MAL-051,liu2012distributed,lopes2008diffusion,tu2011mobile,cattivelli2011modeling}. 
 Unfortunately, incorporating spatial networks of interactions into the biochemical reaction system greatly increases its complexity and renders efficient modeling difficult. 

Recently, first attempts have been made to investigate the effect of cell-to-cell communication on the fidelity of environmental sensing \cite{fancher2017fundamental,camley2017cell}. For example, Fancher and Mugler \cite{fancher2017fundamental} have shown theoretically that communication can significantly enhance the accuracy at which external signaling gradients can be resolved. While the theory presented in this study revealed important new insights, it applies only to static environmental signals. Moreover, their analysis was restricted to simple interaction topologies consistent with diffusion in three dimensions.

In this letter, we develop a framework to study the sensing accuracy of dynamic environmental signals in networks of cells with arbitrary interaction topology. We introduce a formalism that rigorously captures molecular fluctuations in the sensing and communication circuitry. Based on this framework, we study the interplay between dynamics and interaction topology and their effect on environmental sensing with Poissonian and super-Poissonian statistics.  We first investigate this interplay with some biologically inspired case-studies (bacterial-like fully-connected population and epithelial-like hexagonal lattice). Then to provide better insight into the important properties of networks for effective communication, we employ simple models of random networks to study dependence of the quality of estimation on local and global network measures (e.g. degree, clustering coefficient, and connectedness).
 
We consider a population of $N$ cells exposed to an environmental signal $Z(t)$. The signal $Z(t)$ could, for instance, be the concentration of a signaling molecule acting upon the population. We model $Z(t)$ as a one dimensional birth-death process:
\begin{equation}
\emptyset \xrightharpoonup{~\rho~} Z \xrightharpoonup{~\varphi~} \emptyset
\label{eq:Z_reac}
\end{equation}
with birth-rate $\rho$ and death-rate $\varphi$.  We next assume that the cell $i$ needs to estimate the signal up to a proportionality constant, i.e. $\gamma Z(t)$. For simplicity, we consider the cell being able to sense the external signal $Z(t)$ via a single catalytic reaction:
\begin{equation}
Z \xrightharpoonup{~\gamma c_M~} Z + M^{(i)}.
\label{reac:1st_sensor}
\end{equation}
Here, $\gamma c_M$ is a constant rate where $c_M$ and $\gamma$ are the sensor rate and cell's enhancement factor (i.e. the proportionality constant of the estimation), respectively. Whenever a sensor reaction happens, a molecule $M^{(i)}$ is produced inside cell $i$, which is then available for downstream processing.
Note that the firing times of the sensor reaction provide noisy and indirect measurements of $Z(t)$ to the cell and the enhancement factor $\gamma$ controls the informativeness of these measurements. For large $\gamma$, for instance, the sensor reaction fires more frequently and thus follows the abundance of $Z(t)$ more closely. 

We next introduce an estimator circuit, which processes the molecules $M^{(i)}$ produced by the sensor to construct an estimate of $\gamma Z(t)$. In this study we focus on a linear birth-and-death process
\begin{equation}
\emptyset \xrightharpoonup{~~\eta~~} M^{(i)} \xrightharpoonup{~~\zeta~~} \emptyset,
\label{reac:estm_snglcell}
\end{equation}
with $\eta$ and $\zeta$ as constant rates. In the absence of the sensor reaction, the abundance of the molecule $M^{(i)}(t)$ exhibits Poissonian fluctuations at stationarity and we thus refer to this estimator as a \textit{Poissonian estimator}.

Note that if $\eta=\gamma \rho$ and $\zeta=\varphi + c_M$, the abundance of the molecule $M^{(i)}(t)$ yields an approximation of the optimal Bayesian estimator of $\gamma Z(t)$ (see Ref. \cite{zechner2016molecular} for the details) and we adopt this near optimal choice of parameters in the present work.
Similar models of concentration sensing have been widely used in the literature \cite{fancher2017fundamental, camley2017cell}.

We next extend this model to allow cells to exchange information with other cells in their neighborhood. In particular, we consider the case where estimator molecules can diffuse back and forth between two neighboring cells at a fixed rate constant, i.e., 
\begin{equation}
M^{(j)} \xrightleftharpoons[~~\alpha_{ji}~~]{\alpha_{ij}} M^{(i)}
\label{eq:intrction_reaction}
\end{equation}
where $\alpha_{ij}$ defines the rate of transportation from cell $j$ to $i$. Note that $\alpha_{ij}=0$ if cells $i$ and $j$ do not interact with each other. Throughout this letter we consider symmetric interactions, i.e. $\alpha_{ij}=\alpha_{ji}=\alpha$ for every connected $i$ and $j$. Such interactions will equalize cell-to-cell differences in concentration, causing a net flux of estimator molecules from cells containing more towards neighboring cells with fewer estimator molecules.

Considering each cell's environment to be well-mixed, we can describe the stochastic time-evolution of the environmental signal $Z(t)$ and each cell's estimator molecule $M^{(i)}(t)$ by a continuous-time Markov chain \cite{van1992stochastic, anderson2011continuous}. In particular, we resort to a counting process formalism, where $Z(t)$ and $M^{(i)}(t)$ are described by a system of stochastic integral equations which have independent unit Poisson processes counting the occurrences of the reactions \ref{eq:Z_reac}--~\ref{eq:intrction_reaction} (see appendix \ref{sec:celcmn}).

To assess the bias and accuracy of each cells estimator, we define the Mean Error (ME) 
$
	\mathbb{E}\left[e_{i}(t)\right] = \mathbb{E}\left[ Z(t)- \nicefrac{M^{(i)}(t)}{\gamma} \right] 
$
and the Mean Squared Error (MSE) 
$
	\mathbb{E}\left[e^2_{i}(t)\right] = \mathbb{E}\left[ \left(Z(t)- \nicefrac{M^{(i)}(t)}{\gamma} \right)^2 \right]
$
, respectively. Differential equations for the ME and MSE can be elegantly derived from (\ref{eq:rtcm_Z}) and (\ref{eq:rtcm_M}) using Ito's lemma for counting processes (see appendix \ref{sec:celcmn}).

\begin{figure}
\includegraphics[width=0.8\linewidth]{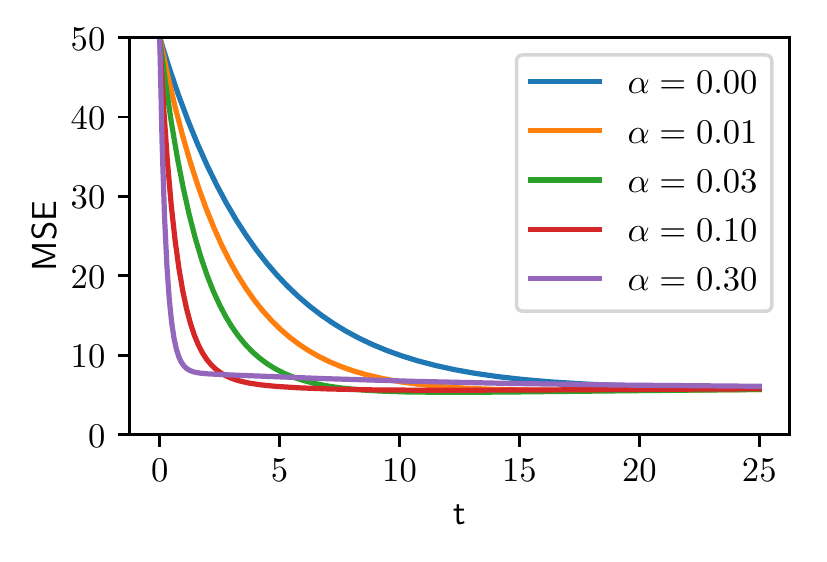}
\caption{\label{fig:MSE_dyn_vs} Dynamics of MSE of a Poissonian estimator in a fully-connected network with 5 neighbors and different coupling strength $\alpha$ with the fully-connected network of cells with following set of parameters: $\rho=c_M=0.1$, $\varphi=0.01$, and $\gamma=2$.}
\end{figure}

We first consider a mean field configuration where a population of cells with size $N$ is described as a fully connected network of Poissonian estimators. This interaction topology applies, for instance, to bacterial populations in which each cell can communicate with all the others by secreting fast diffusing molecules into the surrounding media. 
The dynamics of the MSE can be determined by numerical integration of its governing equation as described in appendix \ref{sec:celcmn}. As depicted in Fig. \ref{fig:MSE_dyn_vs}, by increasing the coupling strength $\alpha$, the MSE converges faster to its minimum. 
We show in appendix \ref{sec:eig_full} that for sufficiently large $n$, the convergence rate is approximately given by $\lambda=-2\left(\varphi+c_M+\left(n+1\right)\alpha\right)$, which shows that increasing both $n$ and $\alpha$ will lead to faster convergence of the MSE.

While our analysis shows that communication can boost the convergence rate of the MSE, it does not affect its steady-state value. For a simple fully connected network, for instance, one can prove that 
\begin{align}
&\mathbb{E}[e_i^2]=\frac{\rho}{\gamma\varphi} + \frac{\rho}{c_M+\varphi}\\
&\mathbb{E}[e_i e_j]=\frac{\rho}{c_M+\varphi} \quad j\neq i
\end{align}
which show that both $\mathbb{E}[e_i^2]$ and $\mathbb{E}[e_i e_j]$ are independent of the coupling strength as well as the number of neighbors, and we verified this analytical result using exact stochastic simulations \cite{gillespie2007stochastic} of the considered system (see appendix \ref{sec:flly_cnnctd}).
Note that $\mathbb{E}[e_i^2]$ and $\mathbb{E}[e_i e_j]$ are merely related except a shift by $ \nicefrac{\rho}{\gamma\varphi}$ (i.e. $\mathbb{E}\bigl[(M^{(i)})^2\bigr]-\mathbb{E}[M^{(i)} M^{(j)}]=\frac{\rho}{\gamma\varphi}$). 
In appendix \ref{sec:flly_cnnctd}, we also show that this relation causes the cancellation of the coupling terms from the steady state equations.
This result is generally due to the fact that the MSE is fundamentally bounded by the intrinsic Poissonian fluctuations of the estimator, which cannot be overcome by diffusive transport (i.e. spatial averaging).
This behavior has been seen in the context of gene expression  \cite{erdmann2009role,sokolowski2015optimizing}, but it is new to dynamical signal sensing. In summary, our analysis shows that communication does not affect the steady-state MSE of Poissonian estimators, although it allows cells to reach this steady-state more quickly. This could play an important role during cell fate determination, where cells have to make decisions upon external cues within a limited amount of time.

Having established that communication cannot enhance the steady-state fidelity of environmental sensing in Poissonian estimators, we next consider the case where the estimator circuit exhibits \textit{super-Poissonian} statistics. Super-Poissonian statistics can arise from additional chemical steps in the estimator's dynamics or due to cell-to-cell differences in process parameters. The latter, also known as \textit{extrinsic variability} \cite{Elowitz1183}, has been shown to often be the dominant source of variability in biochemical networks.

We describe a super-Poissonian estimator by introducing a random mismatch between the birth rate of the estimator in cell $i$ and the true birth rate $\eta=\gamma\rho$ of the environmental signal. More precisely, we equip each cell $i$ with a different birth rate $\eta^{(i)}=\gamma\rho^{(i)}=\gamma \bigl(\rho+\Delta \rho^{(i)}\bigr)$ where $\Delta \rho^{(i)}$ for all $i=1, \ldots, N$ are uncorrelated, zero-mean random variables (i.e., $\mathbb{E}[ \Delta \rho^{(i)} ]= \mathbb{E}[ \Delta \rho^{(i)} \Delta \rho^{(j)} ]=0$) with variance $\mathbb{E}[ (\Delta \rho^{(i)})^2 ]=\sigma^2$.

\begin{figure}
\begin{subfigure}{.44\linewidth}
\includegraphics[width=1\linewidth]{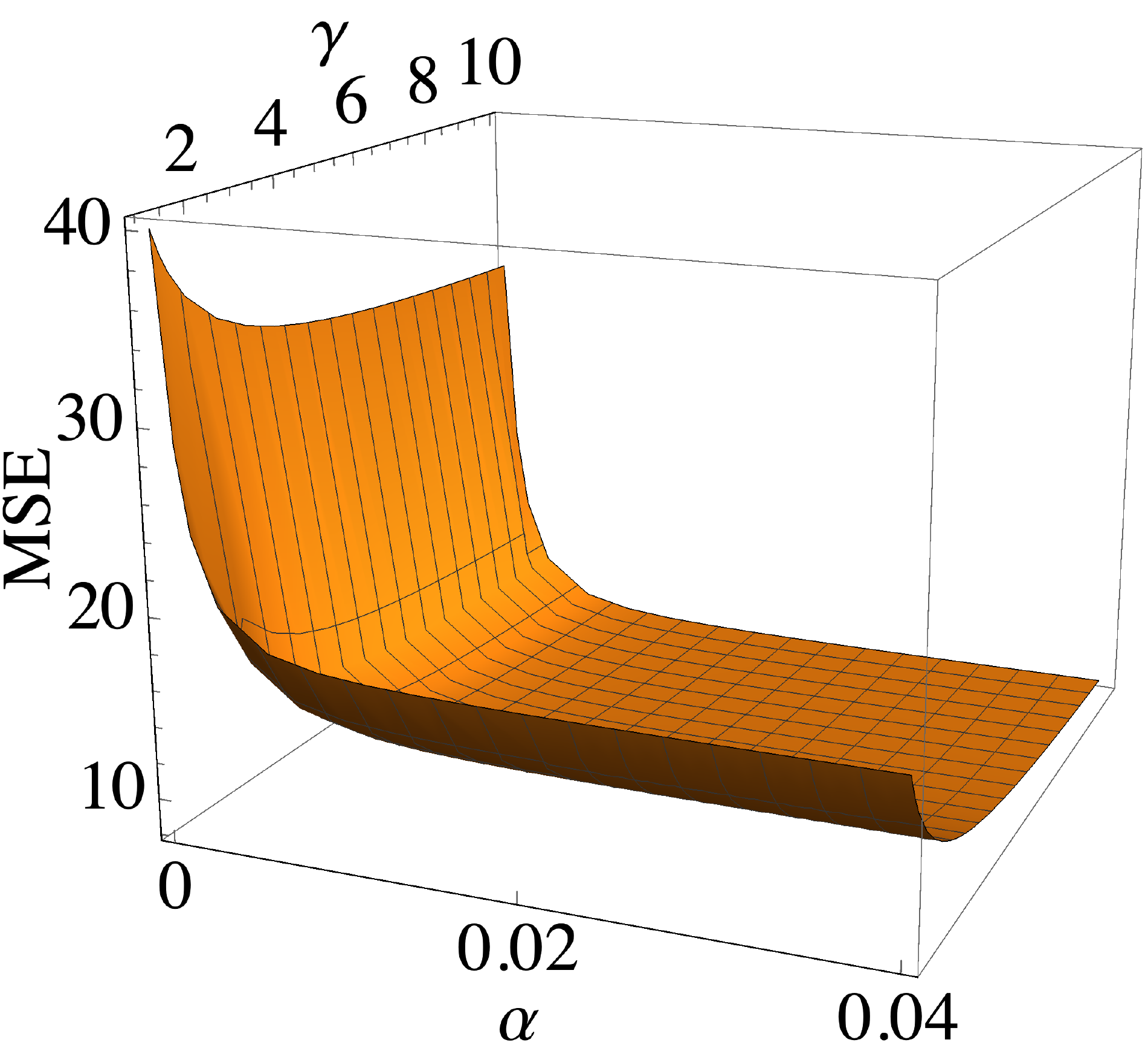}
\caption{}
\label{fig:MSE_vs_ga}
\end{subfigure}
\begin{subfigure}{.46\linewidth}
\includegraphics[width=1\linewidth]{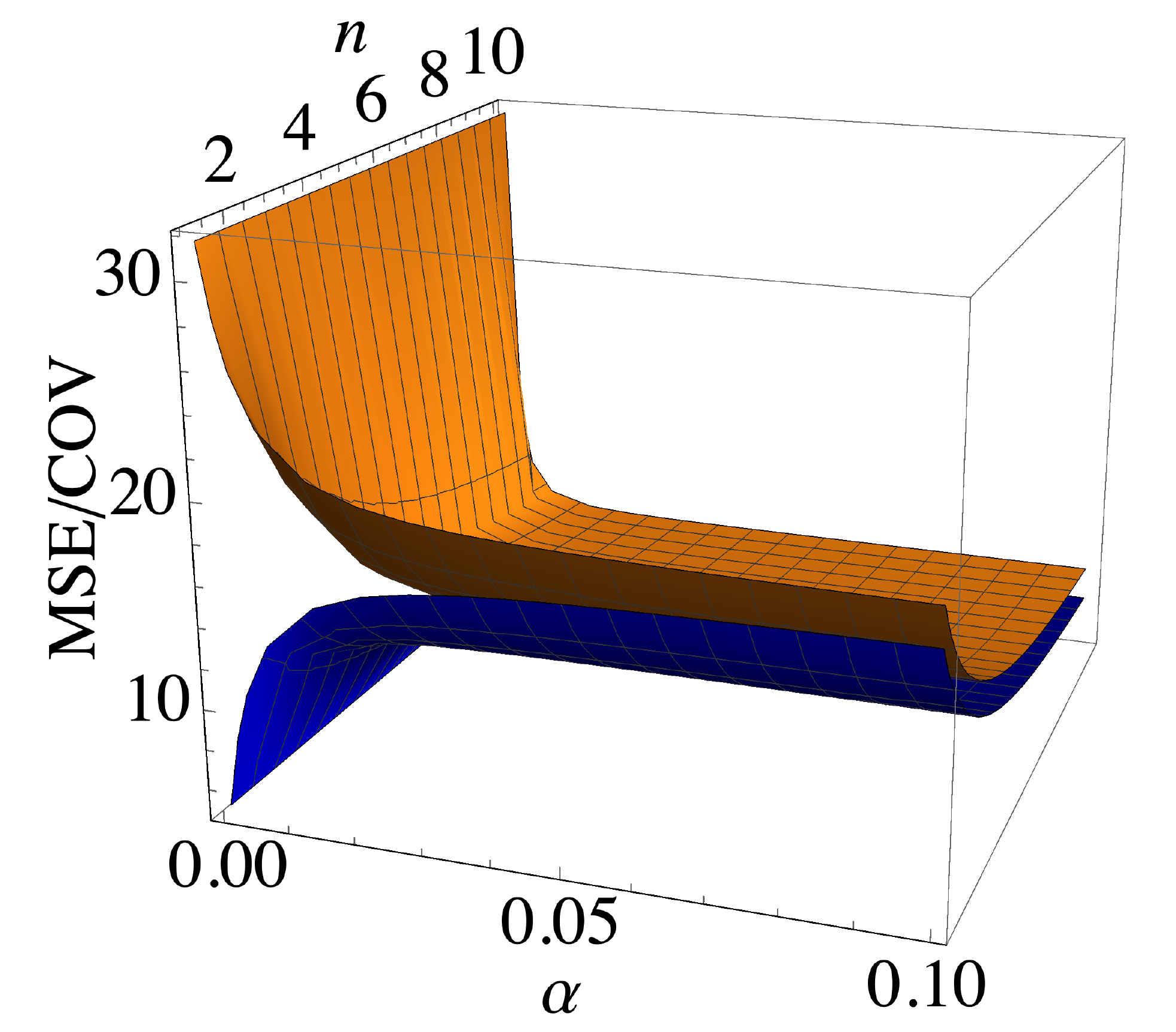}
\caption{}
\label{fig:MSECOV_vs_na}
\end{subfigure}
\caption{\textbf{(a)} MSE vs. coupling strength $\alpha$ and cell's enhancement factor $\gamma$ in the fully-connected network of cells with following set of parameters: $\rho=0.1$, $\varphi=c_M=0.01$, $n=10$ and $\sigma^2=0.01$ \textbf{(b)} MSE (orange surface) and covariance of errors between two cells (blue surface) vs. number of neighbors $n$ and coupling strength $\alpha$ in the fully-connected network of super-Possonian estimators with $\gamma=6$.}
\end{figure}

We now consider the case of fully-connected networks of super-Poissonian estimators and analyze their MSE at steady-state. In particular, we compare how the quality of estimation improves by (I) increasing the coupling strength $\alpha$ and (II) increasing the enhancement factor $\gamma$, which in turn decreases the noise in the sensor and estimator reactions.
Fig. \ref{fig:MSE_vs_ga} shows the MSE of a cell in the population for different values of $\alpha$ and $\gamma$ for a specific set of parameters, but the results don't change qualitatively over a broad range of relevant parameters.
One can see in Fig. \ref{fig:MSE_vs_ga} that even very small values of the coupling strength $\alpha$ can decrease the MSE very significantly, much stronger than increasing the enhancement factor of the estimation. 
It should be noted that in Fig. \ref{fig:MSE_vs_ga}, even when $\gamma \to \infty$ while $\alpha=0$, the MSE will remain larger than in the case where $\gamma=1$ and $\alpha = 0.002$.
These results indicate that cell-to-cell communication is more beneficial for improving the quality of estimation than expending energy in producing more copies of the signal.

\begin{figure*}
\begin{subfigure}{.3\linewidth}
  \centering
\includegraphics[width=\linewidth]{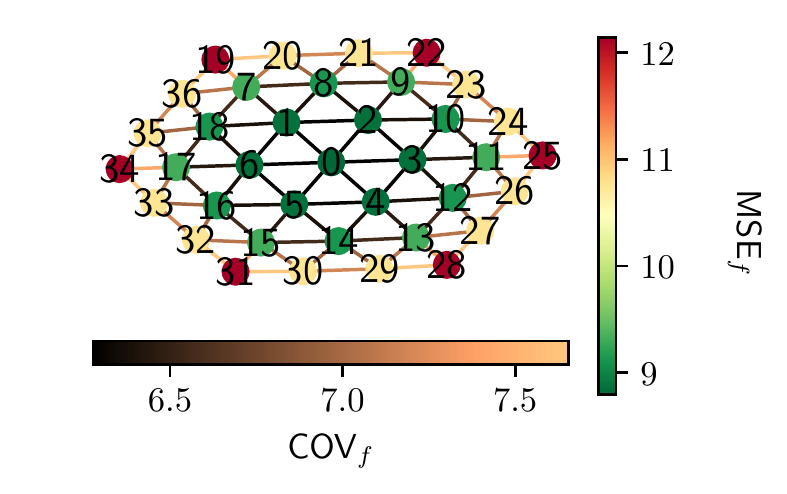}
  \caption{}
  \label{fig:prfct_hex_net}
\end{subfigure}
\begin{subfigure}{.3\linewidth}
  \centering
\includegraphics[width=\linewidth]{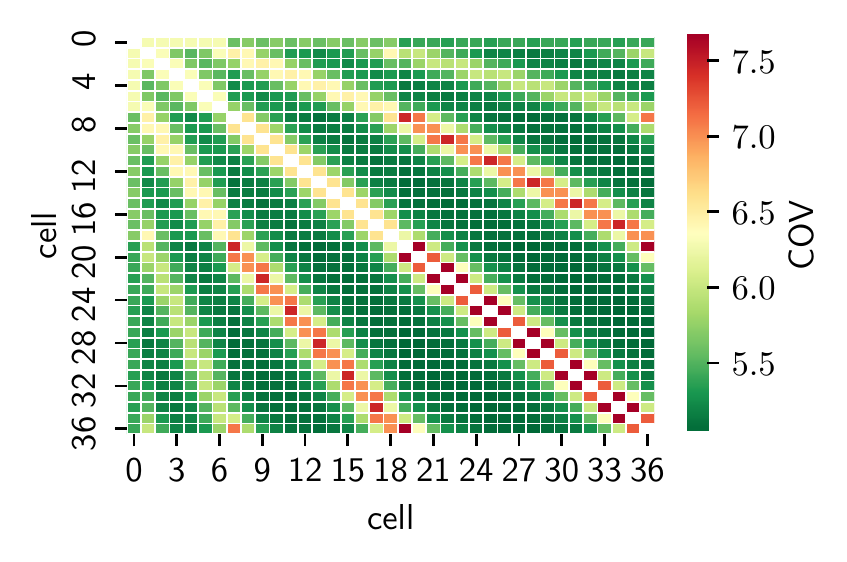}
  \caption{}
  \label{fig:prfct_hex_covmat}
\end{subfigure}
\begin{subfigure}{.3\linewidth}
  \centering
\includegraphics[width=.95\linewidth]{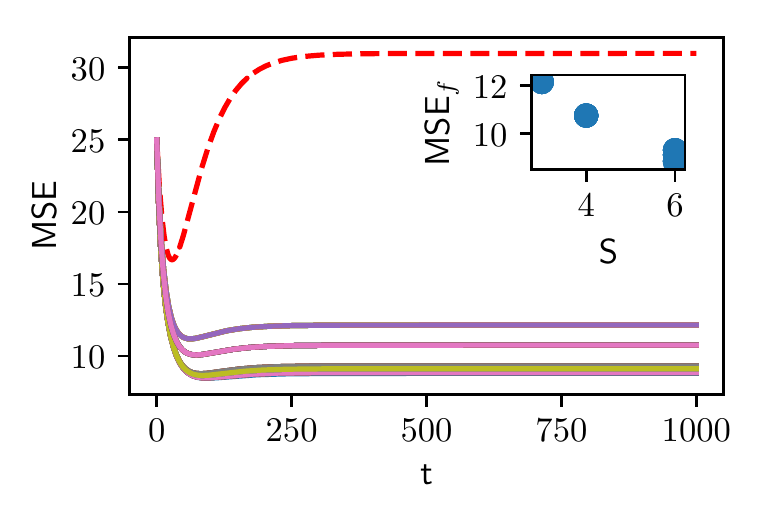}
  \caption{}
  \label{fig:prfct_hex_dyn}
\end{subfigure}
\caption{\textbf{(a)}  Sensing accuracy in a perfect hexagonal lattice at stationary state \textbf{(b)} Covariance between cells in the network at stationary state \textbf{(c)} Dynamics of MSE of cells in the network in comparison with a single cell (red dashed line) starting from the same initial conditions}
\label{fig:prfct_hex}
\end{figure*}

Before further studying the effect of coupling through complex networks, we analyze the effect of coupling between neighbors on the MSE. To this end, we compare the results of fully-connected network with another analytically solvable case: A mean field case in which a hub cell is connected to $n$ neighbors which are not connected to each other. By repeating our analysis for this case (see appendix \ref{sec:star_net}), we find that the MSE of estimation in the hub of the network with $n$ neighbors is always the same as each cell of the fully connected network with $n$ neighbors. Thus, to our surprise, the connection between neighbors (the \emph{local clustering coefficient}) does not affect a cell's MSE of estimation, while it decreases the neighbors' MSEs. In order to test this finding in more complicated cases, we study all intermediate steps of transition from a fully-connected network of a small population to a sparse star-shape network in appendix \ref{sec:exhaustive}. Indeed, even in more complicated scenarios, the MSE at one cell remains unaffected by coupling among its neighboring cells. However, the MSE does depends on the covariances of the errors as one can see in Eq. \ref{eq:MSE}. Accordingly, plotting the MSE and Covariance (COV) together against the number of neighbor $n$ and coupling strength $\alpha$ in either the fully connected or star-shaped network cases (which essentially have the same MSE and COV) can increase our insight into their behaviour. Fig. \ref{fig:MSECOV_vs_na} shows MSE (orange surface) and COV (blue surface) which are correlated to each other when $\alpha$ is fixed and are anti-correlated when $n$ is fixed.

While the effect of nearest neighbors on  the super-Poissonian estimators is evident, the evolution of the MSE also depends on the covariance with the neighbors and the covariance between two cells depends on the covariance of each of them with the neighbors of the other. Therefore, the next nearest neighbors of cells should also play a role in the quality of the estimation. This can be of significant importance in a more realistic case in which the topology of connections is neither fully connected nor star-shaped but lies between these two extremes. We thus study  a biologically inspired case in which cells are epithelial-like, with typically six neighbors placed in a two-dimensional lattice. We start with a perfect 2D hexagonal lattice, and solve the set of equations derived in appendix \ref{sec:celcmn} numerically, to find the MSE of the cells. As one can see in Fig. \ref{fig:prfct_hex}, the quality of estimation as well as the covariances depend on the position of the cells within the lattice. Moreover, the covariance of errors (which is a measure of information transfer between cells) increases when quality of estimation decreases (i.e. the MSE increases). This behaviour indicates that cells with worse estimation receive more information through their available links. In order to further investigate this hypothesis, we use random networks with three different topologies: random spatial, scale-free and small-world networks (see appendix \ref{sec:rndm_graph}).

\begin{figure}
\includegraphics[width=.9\linewidth]{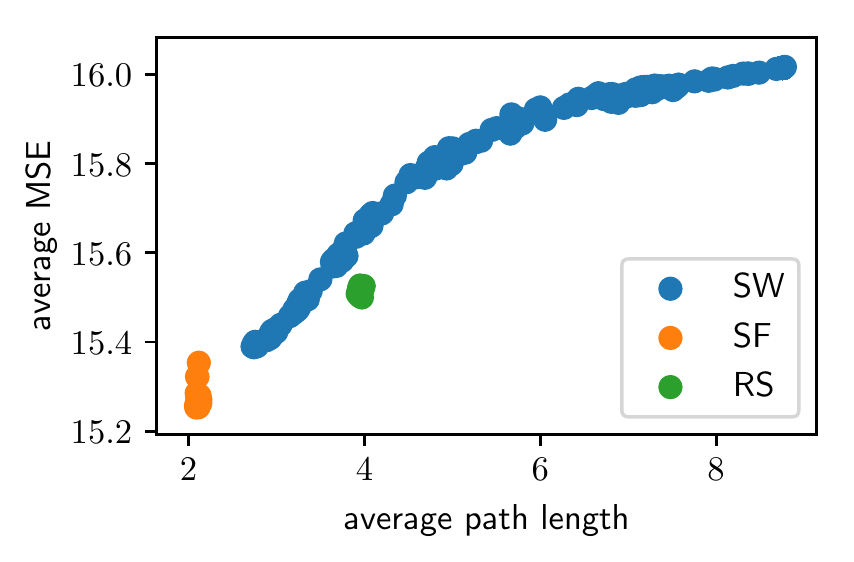}
\caption{\label{fig:MSE_avg} Average MSE vs. average path length for different network topologies: blue circles correspond to SW networks with rewiring probability ranging from 0.0001 to 0.5, orange circles correspond to SF network and green circles show the result of RS networks. In all of these simulations, the mean degree is equal to 6.}
\end{figure}

We start with Random Spatial (RS) networks, relevant for modelling tissues and tissue-like systems. We apply periodic boundary conditions, and cut links with constant probability 0.05, generating 30 different realizations of size 100.  In order to study the effect of coupling in a broader range of degrees, we also generate 30 realizations of Scale-Free (SF) networks with size N=100 by the Barab\'{a}si-Albert model \cite{barabasi1999emergence}.  Finally, since information spreading in complex networks is entangled with its small world property \cite{lu2011small}, we also study realizations of Small-World (SW) topologies for communication by employing the Watts-Strogatz model \cite{watts1998collective}.

We next investigate the relation between a node's MSE and degree in complex networks of super-Poissonian estimators. To do so, we scatter plot the MSE vs. the degree of the node in the three aforementioned topologies, and they all show a strong anti-correlation between the MSE and the degree of the cell (see appendix \ref{sec:rndm_graph}), suggesting that cells with a higher number of neighbors can estimate the environmental signal better. We also study the relation between the covariance of errors of two neighboring cells and their MSE. We scatter plot the covariance of errors on each link vs. the smaller or larger MSE on two ends of the same link . Our results (shown in appendix \ref{sec:rndm_graph}) indicate that there is a strong correlation between the MSE of a cell and covariance of errors with their neighbors. This topology-independent correlation (e.g. RS, SW or SF), confirms our previous hypothesis about the relation between MSE and covariance in the hexagonal lattice and it suggests that cells which have to rely more strongly on a few network links generally achieve worse estimation accuracy than cells which can average over more individual links.

So far we have studied the effect of coupling at the individual cell level, but in the remainder of this article we focus on the effect of coupling on the average MSE over all members of a population. As mentioned before, the MSE of estimation on a node depends not only on its first neighbors, but also on its next neighbors and beyond. However, the effect of more distant neighbors will be less on a given cell as one can see in Fig. \ref{fig:prfct_hex_net} where the effect of boundary cells gets weaker as we move towards the center. Accordingly, the average path length of a network defined as the average number of steps along the shortest paths for all possible pairs of nodes \cite{albert2002statistical} is a strong candidate for the read-out of the effectivness of communication in the network. At a given population size, the cells will be on average more affected by the other cells if the network of interactions has a lower average path length. Therefore, we expect that populations coupled through a network with lower average path length will have better estimation on average. To test this hypothesis, we make SW networks with different rewiring probabilities resulting in different average path lengths and compare their average MSEs. Fig. \ref{fig:MSE_avg} shows that indeed this is the case, depicting the average MSEs of the SW networks with 15 different rewiring probabilities (from 0.0001 to 0.5) and 10 realizations for each rewiring probability. This figure also depicts average MSE of SF and RS networks exhibiting the same behavior as SW networks.

By employing a general framework of signal estimation built on the continuous-time Markov chain formalism, we have been able to thoroughly explore, analytically and \textit{in silico}, the role of communication networks and the sharing of information among otherwise independent estimators.
This general framework can also be applied to similar problems where estimation of a dynamic environmental physical quantity (other than concentration) for a cell population is important. For example, when a cell population is attached to a curved surface, the cells' membrane also acquire curvature. This membrane curvature which is a readout for the surface curvature is sensible via BAR domains \cite{antonny2011mechanisms}.
We find that in our formalism, when the internal dynamics and parameters of the estimators are identical (i.e. the Poissonian statistics limits the accuracy of estimation), then there is no advantage or enhancement of \textit{steady state} estimation that results from \textit{any} communication. However, communication helps them to achieve better estimation at finite times. This transient enhancement can help cells to make decisions in a timely manner as is required, for example, in cell fate determination during development.

When the cells' estimator circuit exhibits super-Poissonian statistics, we find that the situation changes. Communication among estimators now matters at steady state and the interchange of information allows for some manner of collective rectification of the different operating parameters and dynamics. We find that enhanced quality of estimation for a given cell is dominated by the number of neighbors, but, interestingly, we also show that the presence or absence of communication among the first neighbors plays no role in determining the quality. Higher-order neighbors do affect the quality outcome, but in an ever-decreasing manner. If, however, the number of higher order neighbors available to any given estimator grows rapidly with the neighbor-order then their effect may not be ignored out of hand, and indeed, we find that the average quality of estimation across large networks improves as the average path length drops.

We believe that this work represents an important step towards understanding and modelling information processing tasks that take place within a larger, coupled, spatial and topological context. Problems of this type are relevant in biology, where
 epithelial tissues carry out decision processes in the context of the epithelial sheet or in developing mesenchyme where cells flock, migrate, and proliferate in 3d with a dynamically changing neighbor environment.
The diffusive communication considered here is also a necessary first step to considering questions of efficiency and regulation for active, energetically costly communication across such networks.
Distributed swarm computing and more generally distributed computing across a fixed but complex network are also highly relevant design problems where these questions become very important. One might even imagine communication among the scientific research community as a paradigm for these observations, and indeed, while we all have much to communicate, we certainly hope that this communication is in the service of enhancing quality -- or at least achieving consent more quickly!

\bibliography{test} 

\onecolumngrid
\newpage
\appendix

\section{Sensing accuracy in interacting cell communities \label{sec:celcmn}}

As mentioned in the main text, the dynamics of the environmental signal $Z(t)$ and its estimators $M^{(i)}(t)$ can be described by a set of stochastic differential equations which have independent unit Poisson processes counting the occurrences of the reactions, i.e.
\begin{align}
Z(t)=&Z(0)  +  \underbrace{R_{b}^Z \left( \rho t \right)}_\text{birth reaction} -\underbrace{R_{d}^Z \left( \varphi \int _0 ^t  Z(s) \mathrm{d}s  \right )}_\text{death reaction}\label{eq:rtcm_Z} \\
M^{(i)}(t)=& M^{(i)}(0)+\underbrace{R_{s}^{(i)} \left( \gamma c_M  \int _0 ^t  Z(s)  \mathrm{d}s  \right )}_\text{sensor reaction in cell $i$}+  \underbrace{R_{b}^{(i)} \left(\gamma \rho t \right) - R_{d}^{(i)} \left( (\varphi + c_M) \int _0 ^t   M^{(i)}(s)  \mathrm{d}s  \right )}_\text{estimator reactions in cell $i$} \notag \\
&+\sum_{j=1}^N \Biggl[ \underbrace{R_{t}^{i\gets j} \left( \alpha_{ij} \int _0 ^t  M^{(j)}(s)  \mathrm{d}s  \right )}_\text{transport to cell $i$} - \underbrace{R_{t}^{j\gets i} \left( \alpha_{ij}  \int _0 ^t  M^{(i)}(s)  \mathrm{d}s  \right )}_\text{transport from cell $i$}  \Biggr], \label{eq:rtcm_M}
\end{align}
where $ R_{b}^Z,  ~R_{d}^Z,  ~R_{b}^{(i)},~ R_{d}^{(i)}, ~ R_{s}^{(i)}$ and $R_{t}^{j\gets i}$ are independent unit Poisson processes counting the occurrences of the respective reaction. 
A single isolated estimator will therefore follow a stochastic differential equation:
\begin{equation}
\mathrm{d}M^{(i)}(t)=\mathrm{d}R^{(i)}_b(t)-\mathrm{d}R^{(i)}_d(t)+\mathrm{d}R_s^{(i)}(t),
\label{eq:stc_est_evl}
\end{equation}
with $R^{(i)}_b(t)$, $R^{(i)}_d(t)$ and $R^{(i)}_s(t)$ as the reaction counters of the birth, death and sensing reactions, respectively. Note that these can each be decomposed into a predictable part and a zero-mean process such that $R^{(i)}_b(t)=\gamma \rho^{(i)} t + \tilde{R}^{(i)}_b(t)$ and $R^{(i)}_d(t)= (c_M+\varphi) \int_{0}^{t}M^{(i)}(s)\mathrm{d}s + \tilde{R}^{(i)}_d$.
However, in the presence of cell-to-cell communication, one needs to take into account the molecule exchange. Therefore, we define the net flux of the transport reactions as
\begin{align}
\mathrm{d}R_t^{ij}(t) =& \mathrm{d}R^{i\gets j}(t) - \mathrm{d}R^{j\gets i} (t) \notag \\
=& \bigl( \alpha_{ij}M^{(j)}(t)-\alpha_{ji}M^{(i)}(t)\bigr) \mathrm{d}t + \mathrm{d}\tilde{R}_t^{(ij)}(t),
\label{eq:trnfrm_term}
\end{align}
with $\mathrm{d}\tilde{R}_t^{(ij)}(t) = \mathrm{d}\tilde{R}^{i\gets j} (t) + \mathrm{d}\tilde{R}^{j\gets i}(t)$. By adding this transport reaction to the birth-death reactions of estimator $M^{(i)}$ Eq. \ref{eq:stc_est_evl}, it's dynamics would be
\begin{align}
\label{eq:evl_tw_cl}
\mathrm{d}M^{(i)}(t)=\mathrm{d}R_b^{(i)}(t)-\mathrm{d}R_d^{(i)}(t)+\sum_{j} \mathrm{d}R_t^{(ij)}(t)+\mathrm{d}R_s^{(i)}(t)
\end{align}
where $i,j=1,2, ... , N$ are indices and $N$ is the size of the system (i.e. the number of estimators). Note that in practice, these interactions are symmetric and $\alpha_{ij}=\alpha_{ji}$.

Now, by substituting values of $R_t^{(ij)}$, $R_b^{(i)}$ and $R_d^{(i)}$ into Eq. \ref{eq:evl_tw_cl}, the evolution of the estimator in cell $i$ becomes
\begin{align}
\mathrm{d}M^{(i)}(t)=& \bigl[\gamma(\rho+\Delta \rho^{(i)})-(\varphi+c_M)M^{(i)}(t)+\gamma c_M Z(t) + \sum_j \bigl(\alpha_{ij}M^{(j)}(t) -\alpha_{ji}M^{(i)}(t)\bigr) \bigr]\mathrm{d}t\notag\\
&+ \mathrm{d}\tilde{R}_b^{(i)}(t)-\mathrm{d}\tilde{R}_d^{(i)}(t) +\mathrm{d}\tilde{R}_s^{(i)}(t)+\sum_j \mathrm{d}\tilde{R}_t^{(ij)}(t),
\label{eq:net_evl}
\end{align}
which is the stochastic differential equation describing the dynamics of estimator $i$. The third term in this equation stems from the interactions with the neighboring cells and it is commonly known as a \emph{consensus} term \cite{ren2005consensus}. Taking the expectation of Eq. \ref{eq:net_evl} gives:
\begin{align}
\label{eq:cm_estm_evl}
\frac{\mathrm{d}}{\mathrm{d}t}\mathbb{E}[M^{(i)}(t)]=& \gamma\rho -(\varphi+c_M)\mathbb{E}[M^{(i)}(t)]+\gamma c_M \mathbb{E}[Z(t)]+\sum_j\bigl(\alpha_{ij}\mathbb{E}[M^{(j)}(t)] -\alpha_{ji}\mathbb{E}[M^{(i)}(t)]\bigr).
\end{align}
Note that the time-evolution of estimators are coupled to each other and also to the expectation of the environmental signal $\mathbb{E}[Z(t)]$, whose time evolution similarly satisfies:
\begin{align}
\label{eq:Z_evl}
\frac{\mathrm{d}}{\mathrm{d}t}\mathbb{E}[Z(t)]= \rho-\varphi\mathbb{E}[Z(t)].
\end{align}

\begin{figure}
\includegraphics[width=0.4\linewidth]{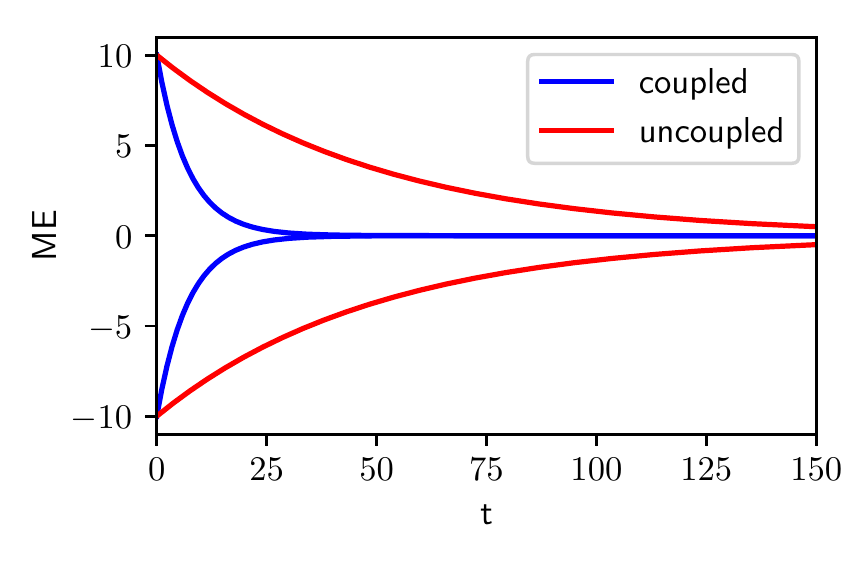}
\caption{\label{fig:2cell} Evolution of expected error for estimators in two cases of coupled (blue curve) and uncoupled (red curve) where parameters are chosen as $c_M=0.01$, $\rho=0.1$, $\varphi=0.01$ and $\alpha=0.04$.}
\end{figure}

\begin{figure}
\begin{subfigure}{.4\linewidth}
  \centering
\includegraphics[width=\linewidth]{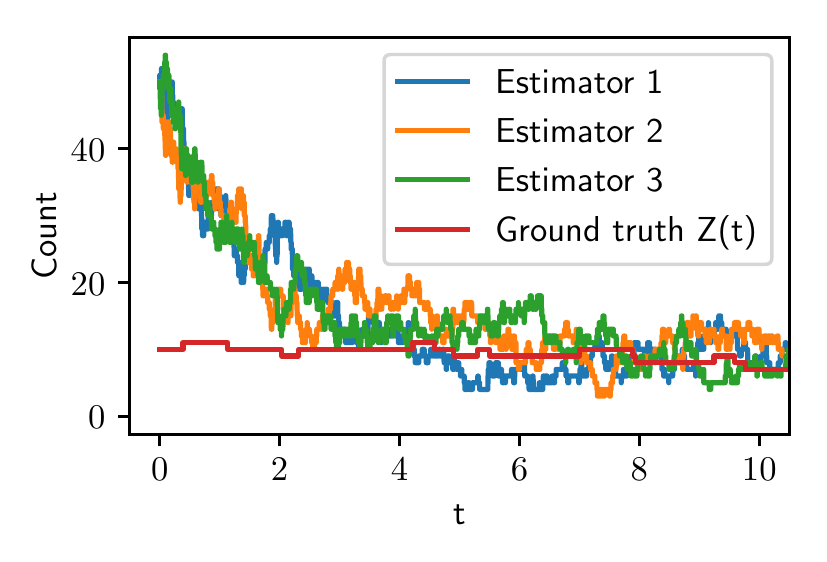}
  \caption{}
  \label{fig:trajs_zoomed}
\end{subfigure}
\begin{subfigure}{.4\linewidth}
  \centering
\includegraphics[width=\linewidth]{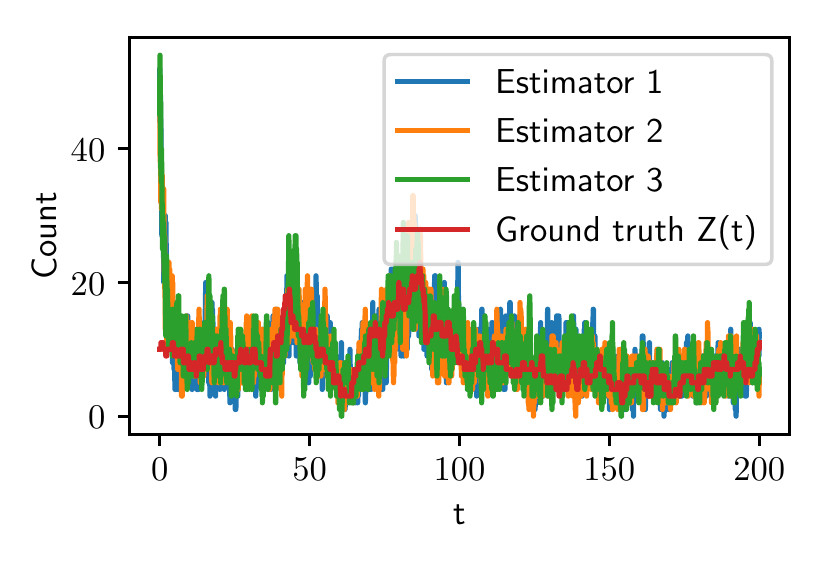}
  \caption{}
  \label{fig:trajs_tst}
\end{subfigure}\\
\begin{subfigure}{.4\linewidth}
  \centering
\includegraphics[width=\linewidth]{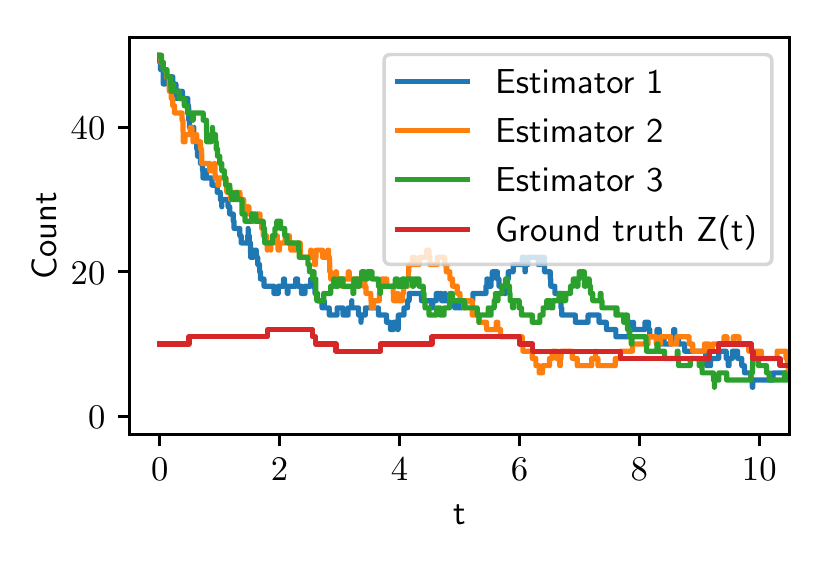}
  \caption{}
  \label{fig:trajs_uncoupled_zoomed}
\end{subfigure}
\begin{subfigure}{.4\linewidth}
  \centering
\includegraphics[width=\linewidth]{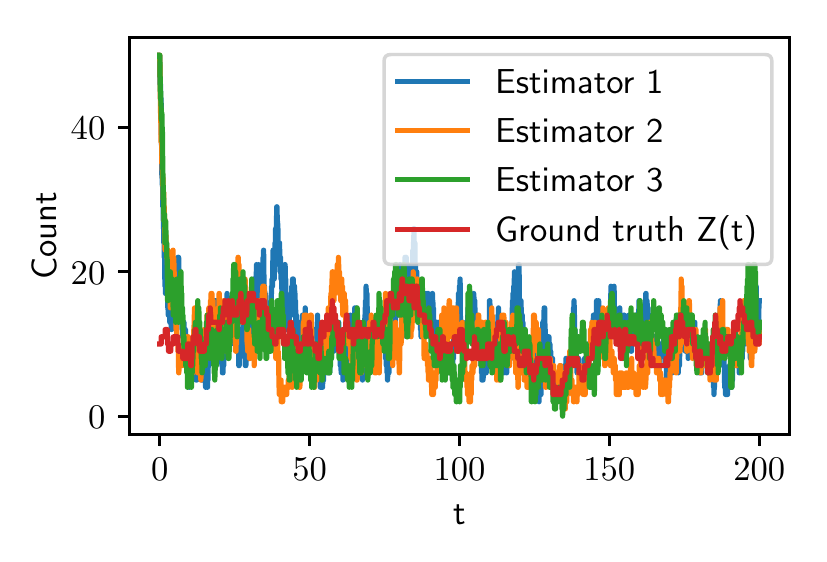}
  \caption{}
  \label{fig:trajs_uncoupled}
\end{subfigure}
\caption{Typical trajectories of the environmental signal and its Poissonian estimators simulated using Gillespie algorithm \cite{gillespie2007stochastic}. \textbf{(a)} Shows the transient part of the estimation in fully-connected estimators and \textbf{(b)} shows the longer trajectories. \textbf{(c)} and \textbf{(d)} also show the transient and longer trajectories in the case of uncoupled estimators. In these simulations the parameters are chosen as $c_M=0.5$, $\rho=1$, $\varphi=0.1$ and $\alpha=0.8$. \label{fig:traj_stoch}}
\end{figure}

It is also easy to show (by subtracting Eq. \ref{eq:Z_evl} from \ref{eq:cm_estm_evl}) that the expected sensing error of cell $i$ (i.e. $\mathbb{E}[e_i(t)]=\mathbb{E}[Z(t)-\frac{1}{\gamma}M^{(i)}(t)]$) is
\begin{align}
	\frac{\mathrm{d}}{\mathrm{d}t}\mathbb{E}[e_i(t)]=&-(\varphi+c_M)\mathbb{E}[e_i(t)] + \sum_j\bigl(\alpha_{ij}\mathbb{E}[e_{j}(t)] -\alpha_{ji}\mathbb{E}[e_{i}(t)]\bigr).
	\label{eq:ME_app}
\end{align}
By simultaneously  solving equations \ref{eq:ME_app} for specific initial conditions, one arrives at the time-evolution of the expected error of the estimation. Typical trajectories of ME for a two cell system is depicted in Fig. \ref{fig:2cell} for two cases of coupled and uncoupled cells. As one can see in this figure, the coupling or communication between cells help them to achieve zero ME much faster than in the uncoupled case.
Moreover, the trajectories of the environmental signal in comparison with three Poissonian estimators in two cases of fully-connected and uncoupled are depicted in Fig. \ref{fig:traj_stoch}. As one can see from comparing Figs. \ref{fig:trajs_zoomed} and \ref{fig:trajs_uncoupled_zoomed} which respectively represent the transient dynamics of the estimators in coupled and uncoupled populations, coupled populations of estimators approach the ground truth signal $Z(t)$ faster.
However, both populations of estimators can estimate not only the steady state value of $Z(t)$, but also resolve its dynamics.

Since the stationary state of the mean error is always zero, the Mean Squared Error (MSE) represents a better measure for quality of the estimation. In order to calculate this, we employ the It\^{o} formula for counting processes, which can be formulated for our special case as following: assume a counting process $X(t)$ has the form $\mathrm{d}X(t)=a \mathrm{d}N(t)$, where $\mathrm{d}N(t)$ is a counting process. Any function of this process $F(X(t))$ will then evolve in time as
\begin{align}
\label{eq:ito}
\mathrm{d}F(X(t))= [F(X(t)+a)-F(X(t))]\mathrm{d}N(t).
\end{align}
For convenience, we separate the transport reaction term in Eq. \ref{eq:trnfrm_term} as $\mathrm{d}R_t^{(i)}=\mathrm{d}R_{in}^{(i)}-\mathrm{d}R_{out}^{(i)}$ in which $\mathrm{d}R_{in}^{(i)}=\sum_j \mathrm{d}R^{i\gets j}(t)$ and $\mathrm{d}R_{out}^{(i)}=\sum_j \mathrm{d}R^{j\gets i}(t)$. By definition we have
\begin{align}
\mathrm{d}e_i(t)=&\mathrm{d}Z(t)-\frac{1}{\gamma}\mathrm{d}M^{(i)}(t)= \notag \\
&\mathrm{d}Z_b(t)-\mathrm{d}Z_d(t)-\frac{1}{\gamma} \bigl(\mathrm{d}R_b^{(i)}(t)-\mathrm{d}R_d^{(i)}(t)  +\mathrm{d}R_{in}^{(i)}(t) -\mathrm{d}R_{out}^{(i)}(t)+\mathrm{d}R^{(i)}_s(t)\bigr)
\label{eq:err_evl}
\end{align}
using Eq. \ref{eq:ito} (the It\^{o} formula) yields
\begin{align}
\mathrm{d}e_i^2(t)=& \bigl(1+2e_i(t)\bigr)\mathrm{d}Z_b(t) +\bigl(1-2e_i(t)\bigr)\mathrm{d}Z_d(t)\notag \\
& +\bigl(\frac{1}{\gamma^2}+\frac{2}{\gamma}e_i(t)\bigr)\bigr[ \mathrm{d}R_d^{(i)}(t) + \mathrm{d}R_{out}^{(i)}(t)  \bigr]+\bigl(\frac{1}{\gamma^2}-\frac{2}{\gamma}e_i(t)\bigr)\bigr[ \mathrm{d}R_b^{(i)}(t) + \mathrm{d}R_{in}^{(i)}(t)+\mathrm{d}R^{(i)}_s\bigr]
\end{align}
Substituting the values of each term and taking the average over realizations, gives the following differential equation for the MSE of the cell $i$:
\begin{align}
\frac{\mathrm{d}}{\mathrm{d}t}\mathbb{E}[e_i^2(t)]=&\rho(1+\frac{1}{\gamma})-2(\varphi+c_M+s_i)\mathbb{E}[e_i^2(t)]-\frac{1}{\gamma}(\varphi+c_M+s_i)\mathbb{E}[e_i(t)]-2\mathbb{E}[\Delta \rho_i e_i(t)] \notag \\
&+\sum_j \alpha_{ij}\bigl(2\mathbb{E}[e_i(t)e_j(t)]-\mathbb{E}[e_j(t)]  \bigr)+\frac{1}{\gamma}\bigl( \varphi(1+\gamma)+2c_M+2s_i\bigr)\mathbb{E}[Z(t)]
\label{eq:MSE}
\end{align}
where $s_i=\sum_{j=1}^N \alpha_{ij}$ is degree of cell $i$ and $\mathbb{E}[\Delta \rho_i e_i(t)]$ is the covariance of cell's uncertainty and its error of estimation. Since the evolution of MSE depends on the covariance of the errors (i.e. $\mathbb{E}[e_i(t)e_j(t)]$), we also must find its time-evolution to obtain a closed set of equations. To this end, we start with the chain rule, i.e., 
\begin{align}
\mathrm{d}\bigl(e_i(t)e_j(t)\bigr)=e_i(t)\mathrm{d}e_j(t)+\mathrm{d}e_i(t)e_j(t)+\Delta e_i(t) \Delta e_j(t).
\label{eq:err_cov}
\end{align}
The third term in Eq. \ref{eq:err_cov} is always zero unless $de_i$ and $de_j$ jump simultaneously. This happens only if: (1) an estimation molecule is exchanged between cell $i$ and $j$, or (2) a birth or death reaction occurs for the $Z$ species. Using this assumption, substitution of Eq. \ref{eq:err_evl} into Eq. \ref{eq:err_cov}, and finally taking the expectation, results in
\begin{align}
\frac{\mathrm{d}}{\mathrm{d}t}\mathbb{E}[e_i(t)e_j(t)]=& \rho - (2\varphi + 2c_M + s_i+s_j)\mathbb{E}[e_i(t)e_j(t)]+\sum_k \bigl(\alpha_{jk}\mathbb{E}[e_i(t)e_k(t)]+\alpha_{ik}\mathbb{E}[e_j(t)e_k(t)] \bigr)\notag \\
&+\alpha_{ij}\bigl(\mathbb{E}[e_j(t)]+\mathbb{E}[e_i(t)]\bigr) +\bigl(\varphi - \frac{2\alpha_{ij}}{\gamma}\bigr) \mathbb{E}[Z(t)]-\mathbb{E}[\Delta \rho_i e_j(t)]-\mathbb{E}[\Delta \rho_i e_j(t)]
\label{eq:cov}
\end{align}
Similarly, one can easily find the time evolution of  $\mathbb{E}[\Delta \rho_i e_i(t)]$ and $\mathbb{E}[\Delta \rho_i e_j(t)]$
\begin{align}
\frac{\mathrm{d}}{\mathrm{d}t}\mathbb{E}[\Delta \rho_i e_i(t)]&= -(\varphi+c_M+s_i)\mathbb{E}[\Delta \rho_i e_i(t)]+\sum_j \alpha_{ij}\mathbb{E}[\Delta \rho_i e_j(t)]  -\mathbb{E}[\Delta \rho_i ^2] \label{eq:co_rho_err}\\
\frac{\mathrm{d}}{\mathrm{d}t}\mathbb{E}[\Delta \rho_i e_j(t)]&= -(\varphi+c_M+s_j)\mathbb{E}[\Delta \rho_i e_j(t)]+\sum_k \alpha_{jk}\mathbb{E}[\Delta \rho_i e_k(t)] \label{eq:co_rho_err2}
\end{align}
Solving Eqs. \ref{eq:MSE}, \ref{eq:cov},\ref{eq:co_rho_err} and \ref{eq:co_rho_err2} jointly with Eq. \ref{eq:ME_app} gives the time evolution of the MSE (and all other variables) for any given initial conditions and set of parameters.

\section{Dynamics of the estimation in fully connected network of Poissonian estimators \label{sec:eig_full}}
In a fully-connected network (i.e. mean-field approximation), one can consider all components to be identical. Accordingly, the set of equations in the previous section will be simpler, especially if one considers Possonian estimators in which $\Delta \rho_i=0$ for every $i$. In such a simple case, only three equations are needed to fully describe the dynamics of the system:
\begin{align}
\frac{\mathrm{d}}{\mathrm{d}t}\mathbb{E}\left[e_i^2(t)\right]=&2\left(\varphi\left(1+\gamma\right)+c_M+n \alpha\right)\frac{\rho}{\gamma\varphi}-2\left(\varphi+c_M+n\alpha\right)\mathbb{E}\left[e_i^2(t)\right]+2n\alpha\mathbb{E}\left[e_i(t)e_j(t)\right]\notag\\
&+\frac{1}{\gamma}\left(\varphi+c_Mn\alpha\left(1+\gamma\right)\right)\mathbb{E}\left[e_i(t)\right]\label{eq:fll_MSE_dyn}\\
\frac{\mathrm{d}}{\mathrm{d}t}\mathbb{E}\left[e_ie_j(t)\right]=&2\rho\left(1-\frac{\alpha}{\varphi\gamma}\right)-2\left(\varphi+c_M+\alpha\right)\mathbb{E}\left[e_ie_j(t)\right]+2\alpha\mathbb{E}\left[e_i^2(t)\right]+2\alpha\mathbb{E}\left[e_i(t)\right]\label{eq:fll_COV_dyn}\\
\frac{\mathrm{d}}{\mathrm{d}t}\mathbb{E}\left[e_i(t)\right]=&-\left(\varphi+c_M\right)\mathbb{E}\left[e_i(t)\right]. \label{eq:fll_ME_dyn}
\end{align}
The dynamics of ME in equation \ref{eq:fll_ME_dyn} is independent of the other two state variable and one can solve it separately, resulting in an exponential decay with exponent $-\left(\varphi+c_M\right)$, i.e. $\mathbb{E}\left[e_i(t)\right]=\mathbb{E}\left[e_i(0)\right] e^{-\left(\varphi+c_M\right)t}$. Therefore, for finding the dependants of transient dynamics of estimation in this case one only needs to consider equations \ref{eq:fll_MSE_dyn} and \ref{eq:fll_COV_dyn}. Since these equations are linear, their exact form can be written as:
\begin{gather}
 \frac{\mathrm{d}}{\mathrm{d}t}
 \begin{bmatrix} \mathbb{E}[e_i^2(t)] \\ \mathbb{E}[e_i(t) e_j(t)] \end{bmatrix}
 =
 \begin{bmatrix} -2(\varphi+c_M+n\alpha) & 2n\alpha \\ 2\alpha & -2(\varphi+c_M+\alpha) \end{bmatrix} 
 \begin{bmatrix} \mathbb{E}[e_i^2(t)] \\ \mathbb{E}[e_i(t) e_j(t)] \end{bmatrix}
 + \begin{bmatrix} C_1+C_2 \mathbb{E}[e_i(t)] \\ C_3+C_4\mathbb{E}[e_i(t)] \end{bmatrix},
\end{gather} 
where $C_1, C_2, C_3$, and $C_4$ are constants. The eigenvalues of the matrix on the rhs are $-2\left(\varphi+c_M\right)$ and $-2\left(\varphi+c_M+\left(n+1\right)\alpha\right)$, and their corresponding eigenvectors are respectively $\hat{MSE}+\hat{COV}$ and $-n\hat{MSE}+\hat{COV}$, where $\hat{MSE}$ and $\hat{COV}$ are the unit-vectors in the direction of MSE and COV. One should note that in the limit of $n \to \infty$, the latter eigenvector lies on the $\hat{MSE}$, and starting from zero COV and ME, only the second eigenvalue determines the dynamics of the system. In this case the system approaches the steady state faster as $n$ or $\alpha$ increase.

\section{Quality of estimation in a fully connected network of Poissonian estimators at steady state \label {sec:flly_cnnctd} }

In this section we study the effect of coupling on quality of estimation in Poissonian estimators. For sake of simplicity, we consider the fully connected case in which all estimators are connected and topologically identical. Accordingly, all cells have no uncertainty about the birth rate of the environmental signal, i.e. $\Delta \rho_i=0$ for $i=1,2, \dots, N$.  
Using this assumption along with the fact that ME vanishes at stationary state, one can rewrite the equations \ref{eq:MSE} and \ref{eq:cov} at stationary as:
\begin{align}
0&= -(\varphi+c_M+n \alpha)\mathbb{E}[e_i^2]+n\alpha\mathbb{E}[e_i e_j]+\frac{1}{\gamma}\bigl(\varphi (1+\gamma)+c_M+n\alpha \bigr)\frac{\rho}{\varphi} \notag\\
0&= -(\varphi+c_M+\alpha)\mathbb{E}[e_i e_j]+\alpha\mathbb{E}[e_i^2]+\bigl(1-\frac{\alpha}{\varphi\gamma} \bigr)\rho \label{eq:st_flcnctd}
\end{align}
where $n=N-1$ is the number of neighbors of each cell and $j\neq i$. This system of algebraic equations is easily solvable, but the solution is surprisingly independent of coupling:
\begin{align}
\mathbb{E}[e_i^2] &= \frac{\rho}{\gamma\varphi} + \frac{\rho}{c_M+\varphi}\notag\\
\mathbb{E}[e_i e_j] &= \frac{\rho}{c_M+\varphi}. \label{eq:sol_MSE}
\end{align}
In order to investigate how the coupling terms (coupling strength $\alpha$ and number of neighbors $n$) drop out, it is beneficial to write Eqs. \ref{eq:st_flcnctd} in matrix form:
\begin{gather}
 \begin{bmatrix} -(\varphi+c_M)-n\alpha & n\alpha \\ \alpha & -(\varphi+c_M)-\alpha \end{bmatrix} 
 \begin{bmatrix} \mathbb{E}[e_i^2] \\ \mathbb{E}[e_i e_j] \end{bmatrix}
 =
 \begin{bmatrix} -\frac{\rho}{\gamma\varphi}(\varphi(1+\gamma)+c_M)-n\alpha \frac{\rho}{\gamma\varphi}
 \\ -\rho+\frac{\alpha}{\gamma\varphi}\rho \end{bmatrix}
\end{gather} 
One can then rewrite this as
\begin{gather}
\left(\begin{bmatrix} -(\varphi+c_M) & 0 \\ 0& -(\varphi+c_M) \end{bmatrix} +
 \begin{bmatrix} n\alpha \\ -\alpha \end{bmatrix} \begin{bmatrix} -1& 1 \end{bmatrix}\right)
 \begin{bmatrix} \mathbb{E}[e_i^2] \\ \mathbb{E}[e_i e_j] \end{bmatrix}
 =
\begin{bmatrix} -\frac{\rho}{\gamma\varphi}(\varphi(1+\gamma)+c_M)
 \\ -\rho \end{bmatrix}
 +
 \begin{bmatrix}
 n\alpha\\
 -\alpha
 \end{bmatrix} \frac{-\rho}{\gamma\varphi}.
\end{gather} 
As is evident in this from, the effect of coupling is separated into the second term on each side of this equation and thus will cancel out if its coefficients on both sides are equal. This implies that $\mathbb{E}[e_i^2] -\mathbb{E}[e_i  e_j]=\nicefrac{\rho}{\gamma\varphi}$ which is indeed true in our case according to \ref{eq:sol_MSE}. This relation can also be rewritten in terms of second order-moment and covariance of estimators:
\begin{align}
\mathbb{E}\bigl[(M^{(i)})^2\bigr]-\mathbb{E}[M^{(i)} M^{(j)}]=\frac{\rho}{\gamma\varphi}=\frac{1}{\gamma^2}\mathbb{E}[M^{(i)}].
\end{align}

\begin{figure}
\includegraphics[width=0.4\linewidth]{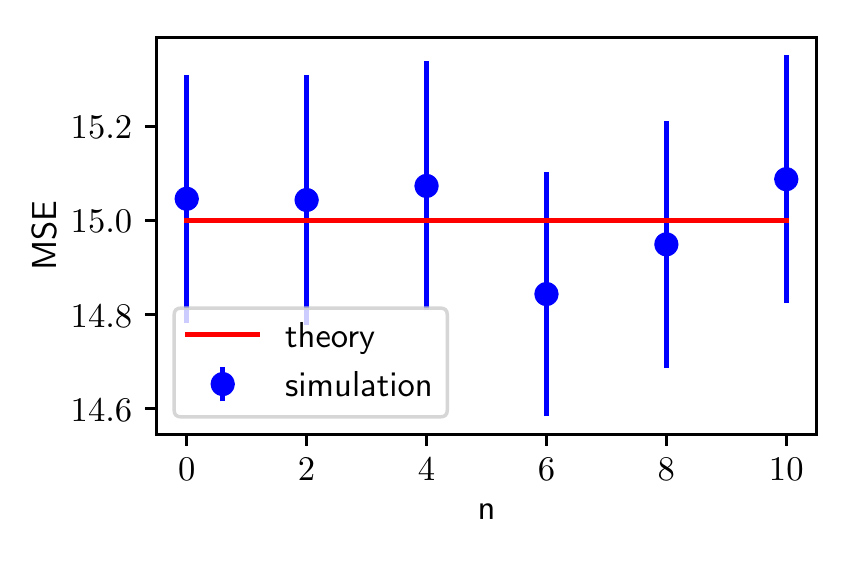}
\caption{\label{fig:flat_MSE} MSE vs. number of neighbors $n=N-1$ in the fully-connected network of Poissonian estimators (i.e. $\sigma^2=0$) with enhancement factor $\gamma=1$. The average (blue circles) and the standard error (shown with the error bars) of MSE are the result of 100000 realizations of stochastic simulations.}
\end{figure}

We also verify the independence of the MSE from the coupling at steady-state by using exact stochastic simulations \cite{gillespie2007stochastic} of a population of estimators with size $N$ communicating across a fully-connected networks. As depicted in Fig. \ref{fig:flat_MSE}, by increasing number of neighbors $n=N-1$, the MSE of estimation does not change. 

\section{Quality of estimation in star-shaped networks \label{sec:star_net}}

In this section, we study another simple yet informative case in which only a hub is connected to the rest of nodes and there are no other connections. Such networks which are known as star-shaped networks, represent a kind of mean-field configuration. In this case, the stationary state version of equations \ref{eq:MSE} and \ref{eq:cov} will be
\begin{align}
0=&-(\varphi+c_M+n\alpha)\mathbb{E}[e_c^2]+n\alpha\mathbb{E}[e_ce_m]-\mathbb{E}[\Delta\rho_c e_c]+\frac{1}{\gamma}\left(\varphi\left(1+\gamma\right)+c_M+n\alpha\right)\frac{\rho}{\varphi}\\
0=&-(\varphi+c_M+\alpha)\mathbb{E}[e_m^2]+\alpha\mathbb{E}[e_ce_m]-\mathbb{E}[\Delta\rho_m e_m]+\frac{1}{\gamma}\left(\varphi\left(1+\gamma\right)+c_M+\alpha\right)\frac{\rho}{\varphi}\\
0=& 2\rho -(2\varphi+2c_M+(n+1)\alpha)\mathbb{E}[e_ce_m]+\alpha\mathbb{E}[e_c^2]+(n-1)\alpha\mathbb{E}[e_me_{m'}]+\alpha\mathbb{E}[e_m^2]-\frac{2\alpha}{\gamma}\frac{\rho}{\varphi}\notag\\
&-\mathbb{E}[\Delta \rho_c e_m]-\mathbb{E}[\Delta \rho_m e_c]\\
0=& -(\varphi+c_M+\alpha)\mathbb{E}[e_me_{m'}]+\alpha\mathbb{E}[e_me_c]+\rho-\mathbb{E}[\Delta \rho_me_{m'}]
\end{align} 
where index $c$ indicates the central cell (hub) while $m$ and $m'$ indicate two distinctive marginal cells which are only connected to the central one. In the case of Poissonian estimators, $\Delta \rho_i$ will be zero for every $i$ and the set of equations will be closed. With this assumption, solving these equations gives us the same MSE independent of coupling similar to the case of fully connected or any other topology. However, in a more realistic case, $\Delta \rho_i$ will not be zero due to the presence of super-Poissonian statistics. In this case one should also consider $[\Delta\rho_i e_j]$ from Eqs. \ref{eq:co_rho_err2} and \ref{eq:co_rho_err} which obey the following equations for different indices
\begin{align}
0=&-(\varphi+c_M+n\alpha)\mathbb{E}[\Delta\rho_c e_c]+n\alpha\mathbb{E}[\Delta\rho_c e_m]-\mathbb{E}[\Delta\rho^2]\\
0=&-(\varphi+c_M+\alpha)\mathbb{E}[\Delta\rho_c e_m]+\alpha\mathbb{E}[\Delta\rho_c e_m]\\
0=& -(\varphi+c_M+n\alpha)\mathbb{E}[\Delta\rho_m e_c]+(n-1)\alpha\mathbb{E}[\Delta\rho_m e_{m'}]+\alpha\mathbb{E}[\Delta\rho_m  e_m]\\
0=& -(\varphi+c_M+\alpha)\mathbb{E}[\Delta\rho_m e_m]+\alpha\mathbb{E}[\Delta\rho_m e_c]-\mathbb{E}[\Delta\rho^2]\\
0=& -(\varphi+c_M+\alpha)\mathbb{E}[\Delta\rho_m e_{m'}]+\alpha\mathbb{E}[\Delta\rho_m e_c].
\end{align} 
Solving this set of algebraic equations shows that the MSE of the hub cell will be equal to the MSE of any of the cells in a fully connected network of the same size.
\begin{figure}
\begin{subfigure}{.24\linewidth}
  \centering
\includegraphics[width=\linewidth]{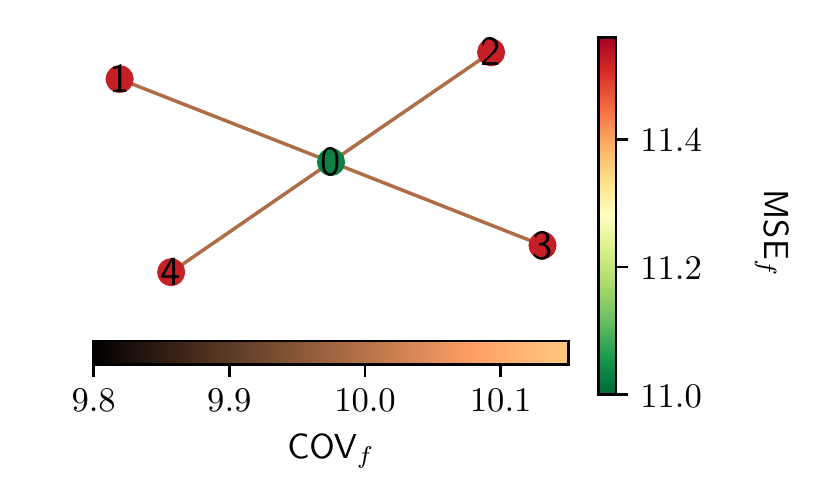}
  \caption{}
  \label{fig:tst0_network}
\end{subfigure}
\begin{subfigure}{.24\linewidth}
  \centering
\includegraphics[width=\linewidth]{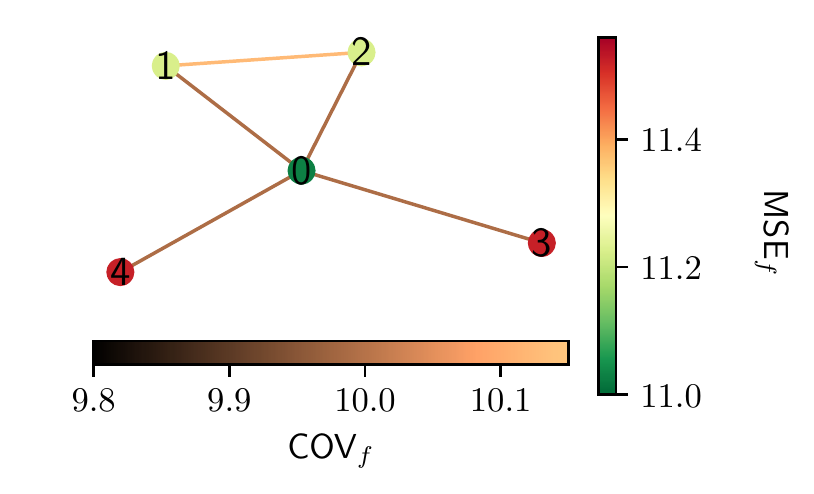}
  \caption{}
  \label{fig:tst1_network}
\end{subfigure}
\begin{subfigure}{.24\linewidth}
  \centering
\includegraphics[width=\linewidth]{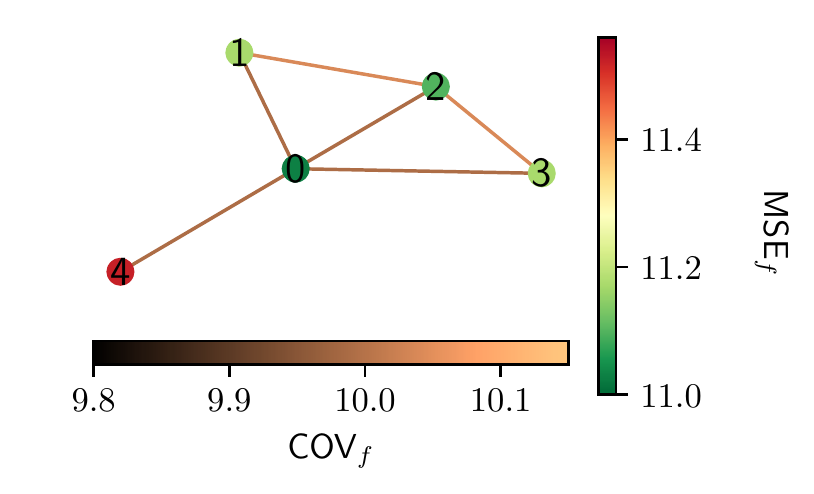}
  \caption{}
  \label{fig:tst2_network}
\end{subfigure}
\begin{subfigure}{.24\linewidth}
  \centering
\includegraphics[width=\linewidth]{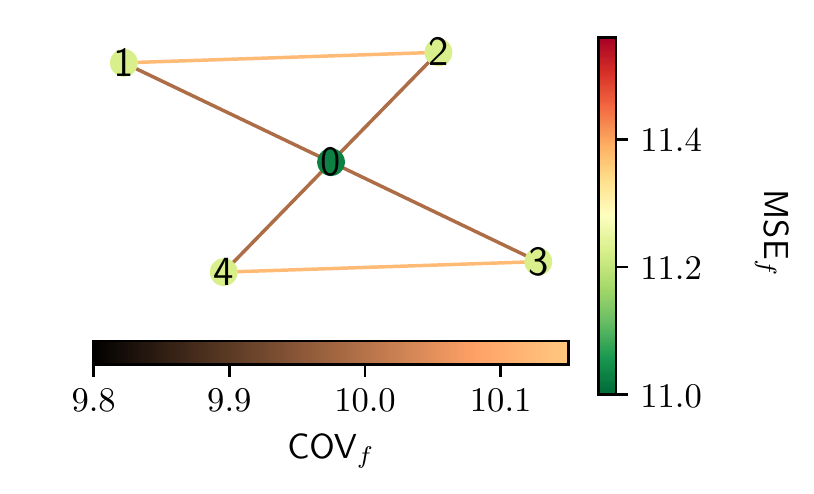}
  \caption{}
  \label{fig:tst3_network}
\end{subfigure}\\
\begin{subfigure}{.24\linewidth}
  \centering
\includegraphics[width=\linewidth]{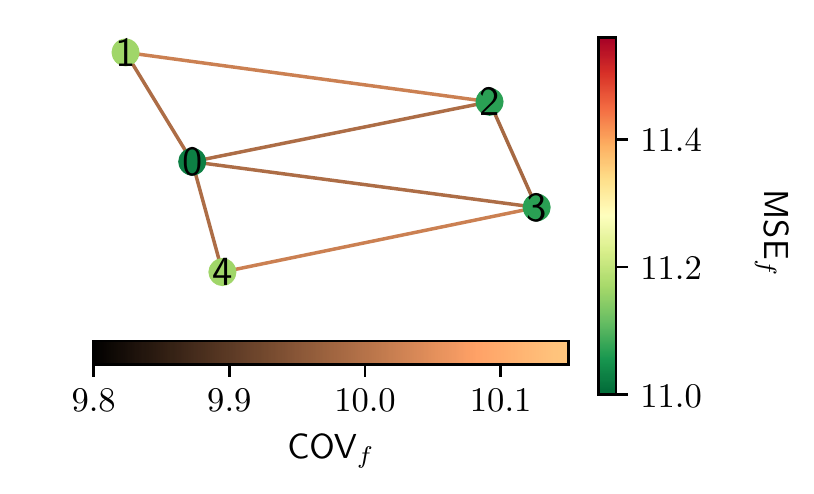}
  \caption{}
  \label{fig:tst4_network}
\end{subfigure}
\begin{subfigure}{.24\linewidth}
  \centering
\includegraphics[width=\linewidth]{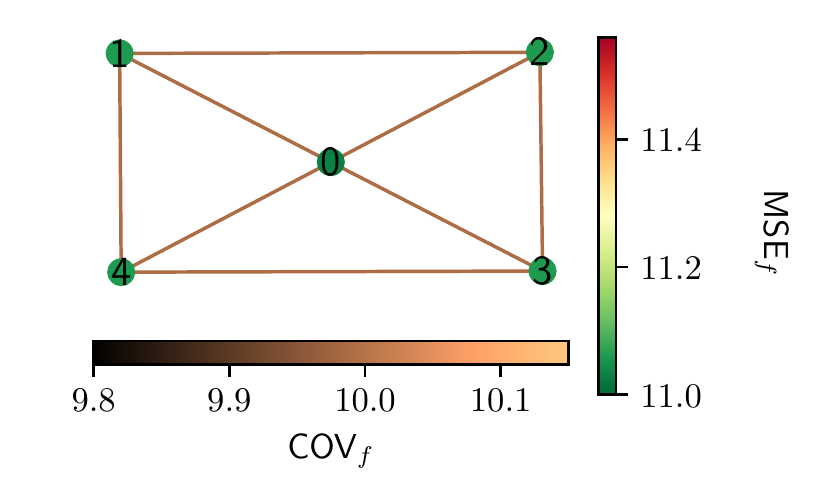}
  \caption{}
  \label{fig:tst5_network}
  \end{subfigure}
  \begin{subfigure}{.24\linewidth}
  \centering
\includegraphics[width=\linewidth]{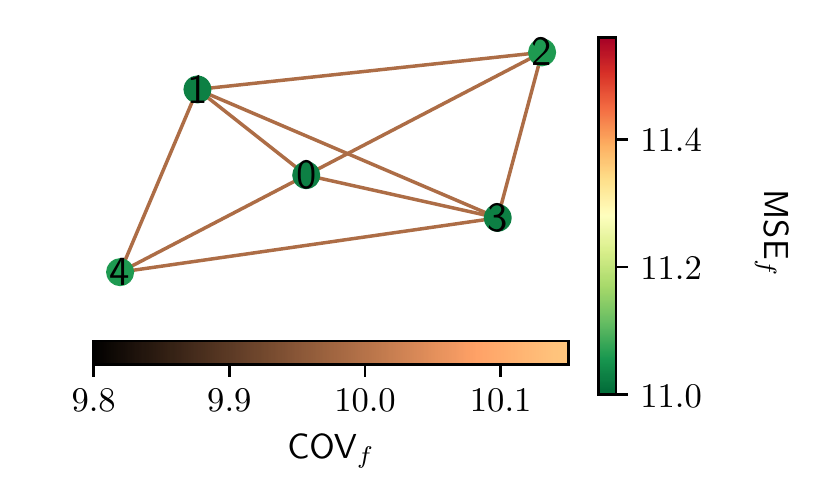}
  \caption{}
  \label{fig:tst6_network}
\end{subfigure}
\begin{subfigure}{.24\linewidth}
  \centering
\includegraphics[width=\linewidth]{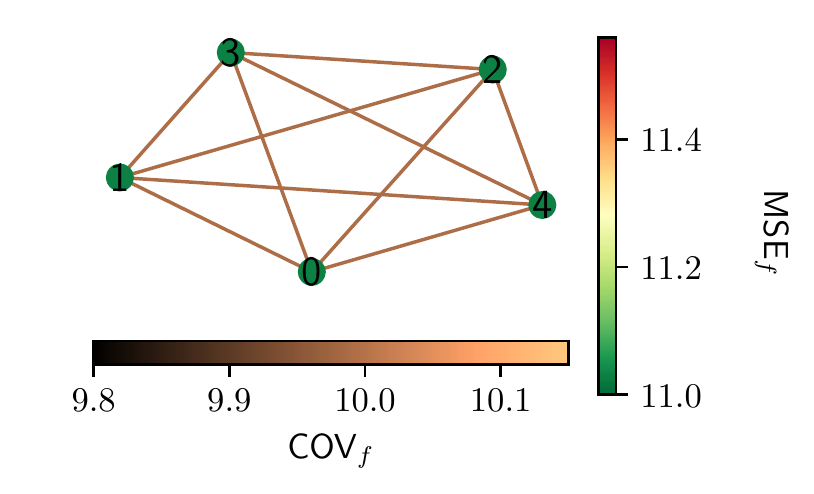}
  \caption{}
  \label{fig:tst7_network}
\end{subfigure}
\caption{ MSE of cells connected through a sparse stare-shaped network \textbf{(a)}, a dense fully-connected network \textbf{(h)} and all intermediate steps \textbf{(b)--(g)}}
\label{fig:exhaustive}
\end{figure}

\section{  MSE of estimation and Transition from star-shaped network to fully connected \label{sec:exhaustive}}

As mentioned in the main text and in Appendix \ref{sec:star_net}, the MSE of the hub of a star-shaped network exactly equals the MSE of a cell in the fully-connected of the same size. In this section, we make all the intermediate steps of transition from a sparse star-shaped network of size $N=5$ to the dense and fully-connected form, and calculate the MSE of the cells as well as the COVs. Fig. \ref{fig:exhaustive} depicts these steps and shows that the MSE of the central cell (number 0) is not affected when links are added among its neighbors. This indicates that MSE does not depend on the local clustering of the system.

\section{Quality of estimation in random networks \label{sec:rndm_graph}}

In this section we study the effect of coupling on the steady-state values of MSE and COV of estimations in random networks. Throughout this section, we use the following set of parameters: $\rho=0.1$, $\phi=0.01$, $c_M=0.01$, $\gamma=1$, $\alpha=0.06$ and $\sigma^2=0.01$. Moreover, for each topology, we generate 30 realizations and in scatter-plots each color corresponds to a realization.

\begin{figure}
\begin{subfigure}{.32\linewidth}
  \centering
\includegraphics[width=\linewidth]{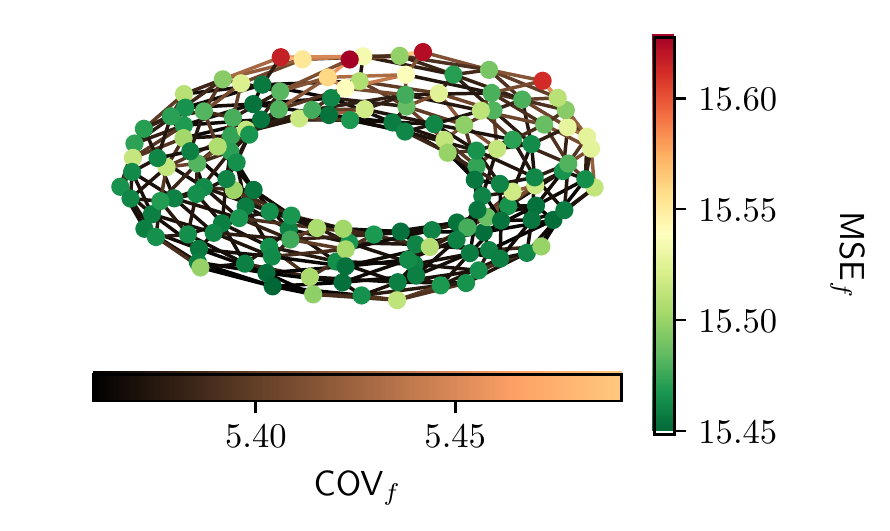}
  \caption{}
  \label{fig:net_RS}
\end{subfigure}
\begin{subfigure}{.32\linewidth}
  \centering
\includegraphics[width=\linewidth]{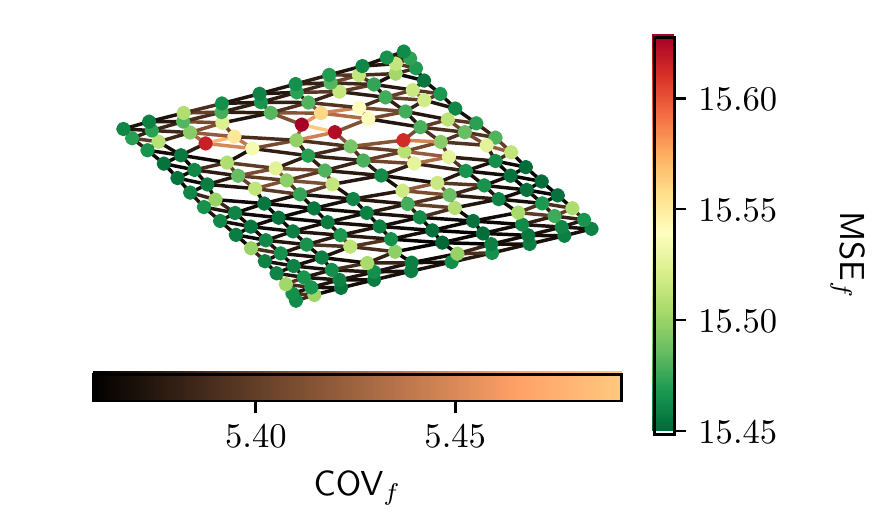}
  \caption{}
  \label{fig:net_uw_RS}
\end{subfigure}\\
\begin{subfigure}{.32\linewidth}
  \centering
\includegraphics[width=\linewidth]{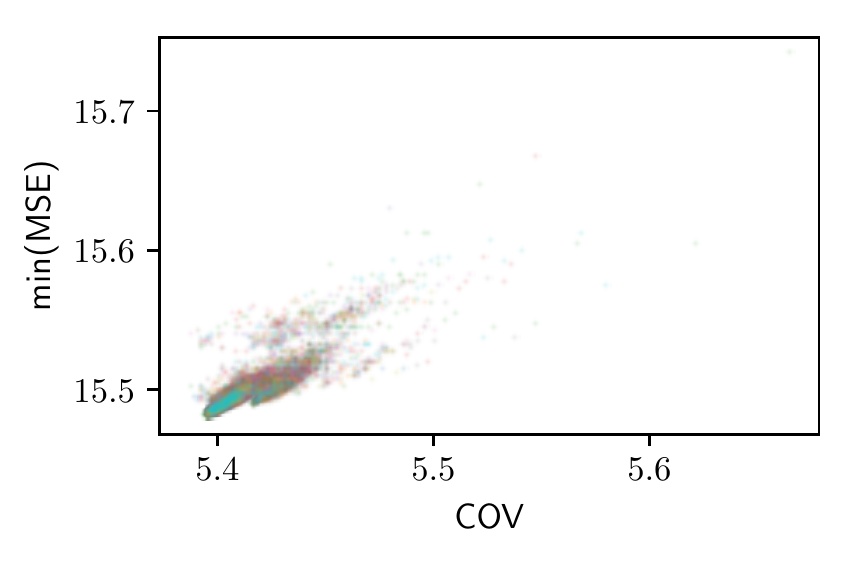}
  \caption{}
  \label{fig:scttr_cov_minMSE_RS}
\end{subfigure}
\begin{subfigure}{.32\linewidth}
  \centering
\includegraphics[width=\linewidth]{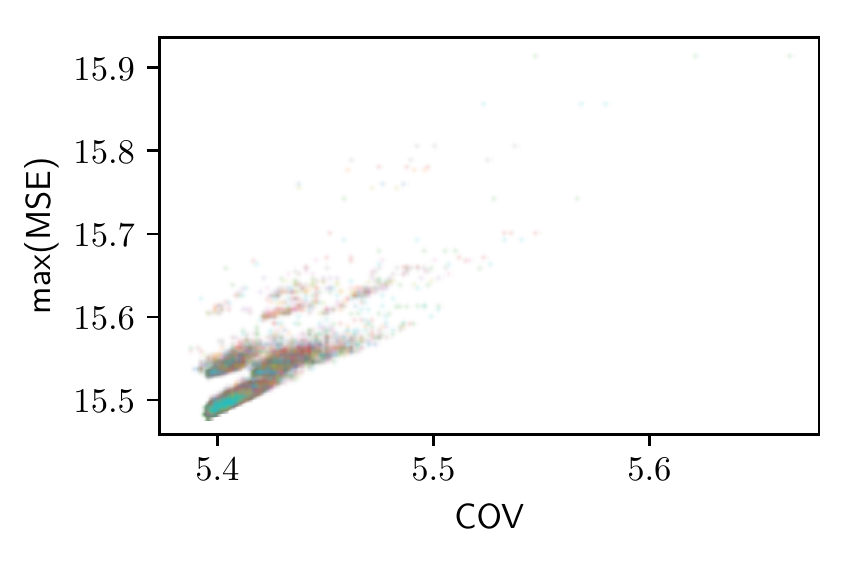}
  \caption{}
  \label{fig:scttr_cov_maxMSE_RS}
\end{subfigure}
\begin{subfigure}{.32\linewidth}
  \centering
\includegraphics[width=\linewidth]{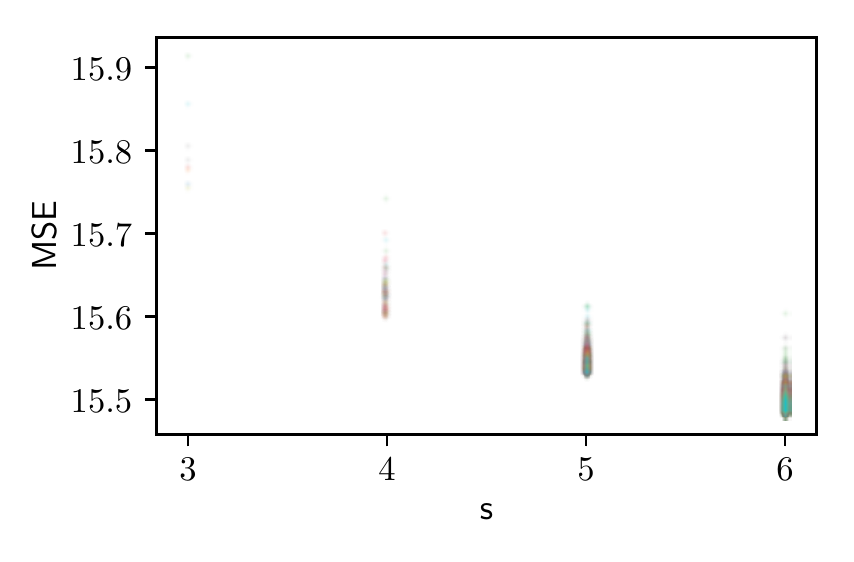}
  \caption{}
  \label{fig:scttr_MSE_deg_RS}
\end{subfigure}
\caption{\textbf{(a)} The MSE and covariance of errors of cells in a typical RS network with real representation. \textbf{(b)} The same network in (a) which is only unwrapped for the sake of visibility \textbf{(c)} Scatter plot of COV of a link vs. the minimum of MSE over the two ends of that link. \textbf{(d)} Scatter plot of COV of a link vs. the maximum of MSE over the two ends of that link. \textbf{(e)} Scatter plot of MSE vs. degree of the nodes s.}
\label{fig:rndm_net_RS}
\end{figure}

We start with the Random Spatial (RS) networks which are generated by connecting both pairs of opposite edges of a hexagonal lattice sheet together and then removing the links with constant probability 0.05. We generated 30 realizations of such random networks with size 100 and the summary of the results for this case is depicted in Fig. \ref{fig:rndm_net_RS}. In Fig. \ref{fig:net_RS} and \ref{fig:net_uw_RS}, a typical RS network is depicted and the MSE of the nodes and COV of errors over the links are coded by different colors. 
Fig. \ref{fig:scttr_cov_minMSE_RS} (and \ref{fig:scttr_cov_maxMSE_RS}) represent the relation between covariance of errors of two neighboring cells and the smaller (and bigger) MSE of estimation between those two cells. Moreover, Fig. \ref{fig:scttr_MSE_deg_RS} shows the scatter plot of MSE of each node vs. its degree (number of neighbors).

\begin{figure}
\begin{subfigure}{.32\linewidth}
  \centering
\includegraphics[width=\linewidth]{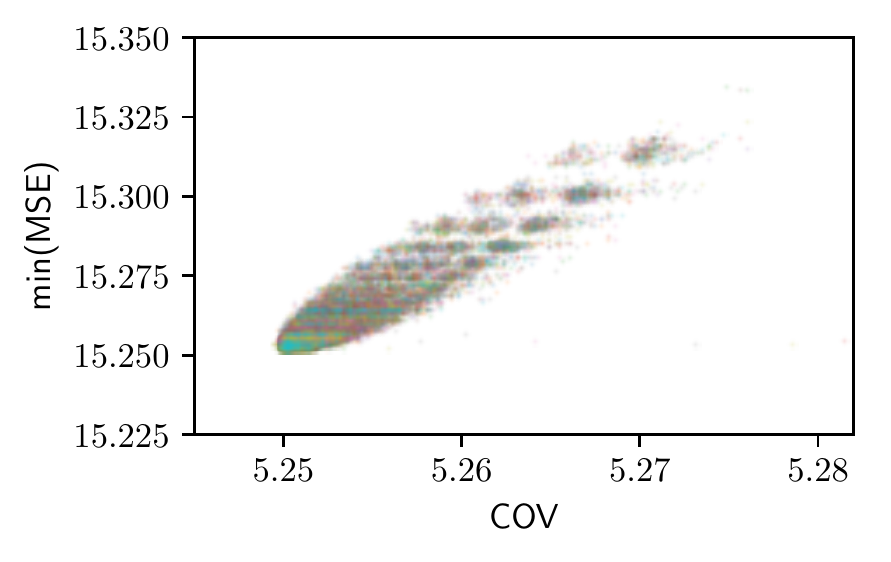}
  \caption{}
  \label{fig:scttr_cov_minMSE_BA}
\end{subfigure}
\begin{subfigure}{.32\linewidth}
  \centering
\includegraphics[width=\linewidth]{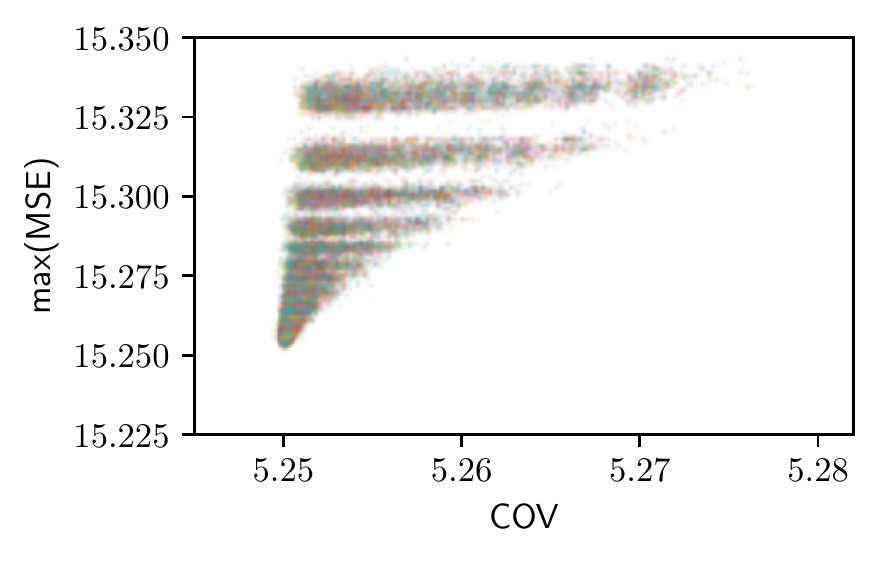}
  \caption{}
  \label{fig:scttr_cov_maxMSE_BA}
\end{subfigure}
\begin{subfigure}{.32\linewidth}
  \centering
\includegraphics[width=\linewidth]{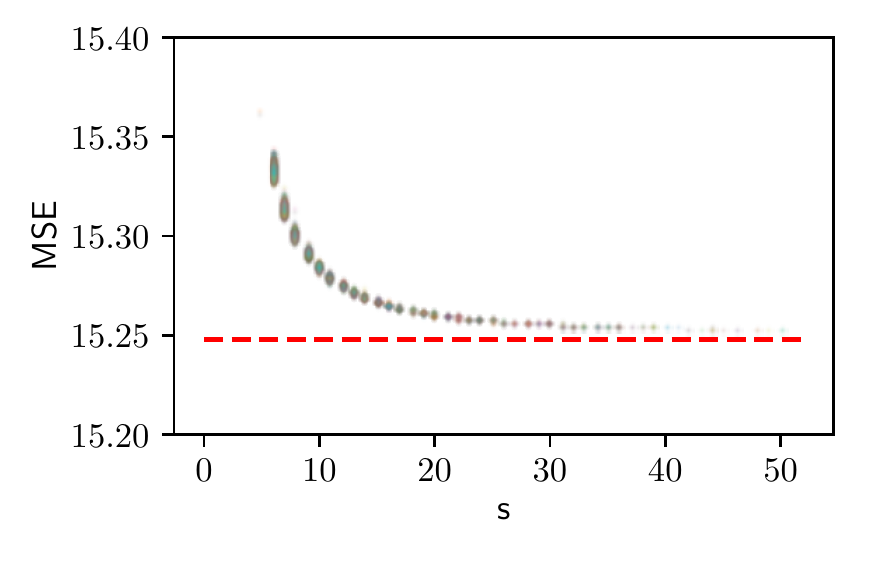}
  \caption{}
  \label{fig:scttr_MSE_deg_BA}
\end{subfigure}
\caption{ \textbf{(a)} Scatter plot of COV of a link vs. the minimum of MSE over the two ends of that link. \textbf{(b)} Scatter plot of COV of a link vs. the maximum of MSE over the two ends of that link. \textbf{(c)} Scatter plot of MSE vs. degree of the nodes s. The red dashed line shows the MSE of a cell in a fully connected network with size 100 and the same set of parameters used in SF network. These figures are result of simulations for 30 different realizations of SF networks with size 100 and mean degree 6.}
\label{fig:rndm_net_BA}
\end{figure}

In order to have a broader range of degrees and see the effect of hubs in the system we study the effect of coupling with Scale-Free (SF) topology. We generate 30 realizations of random SF networks with size 100 and mean degree of 6 using the Barab\'{a}si-Albert model \cite{barabasi1999emergence} and Fig. \ref{fig:rndm_net_BA} depicts our results.
Fig. \ref{fig:scttr_cov_minMSE_BA} shows the scatter plot of COV of errors in two cells connected via a link vs. the minimum of MSE on two ends of that link, and as one can see there is a strong correlation between them. Fig. \ref{fig:scttr_cov_maxMSE_BA} shows the scatter plot of COV on a link vs. the maximum of MSE on two ends of that link. Moreover, Fig. \ref{fig:scttr_MSE_deg_BA} shows the relation between MSE of a node and its degree s. The red dashed line shows the MSE of a cell in a fully connected network with size 100 and the same set of parameters used in SF network. Surprisingly, the hubs of SF networks (although they don't have more than 50 neighbors) have almost the same MSE as the cells in a fully-connected network.

\begin{figure}
\begin{subfigure}{.32\linewidth}
  \centering
\includegraphics[width=\linewidth]{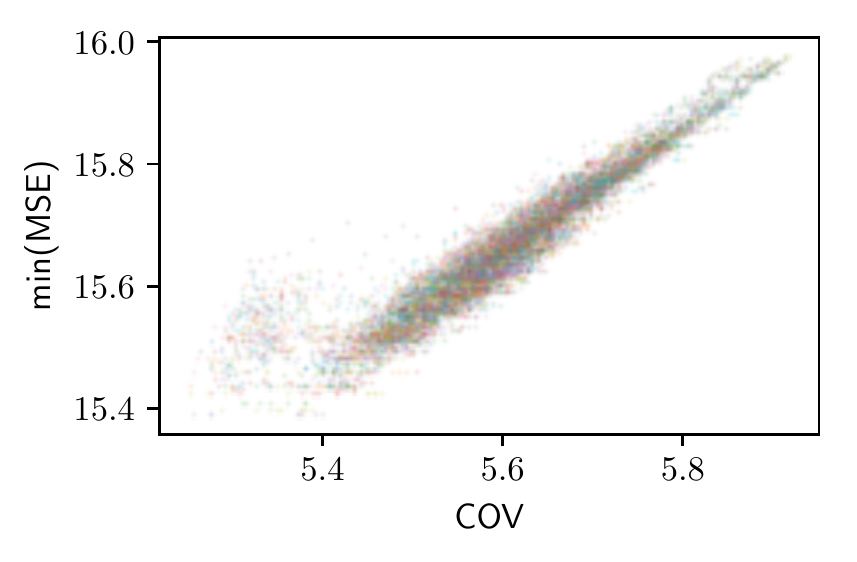}
  \caption{}
  \label{fig:scttr_cov_minMSE_WS}
\end{subfigure}
\begin{subfigure}{.32\linewidth}
  \centering
\includegraphics[width=\linewidth]{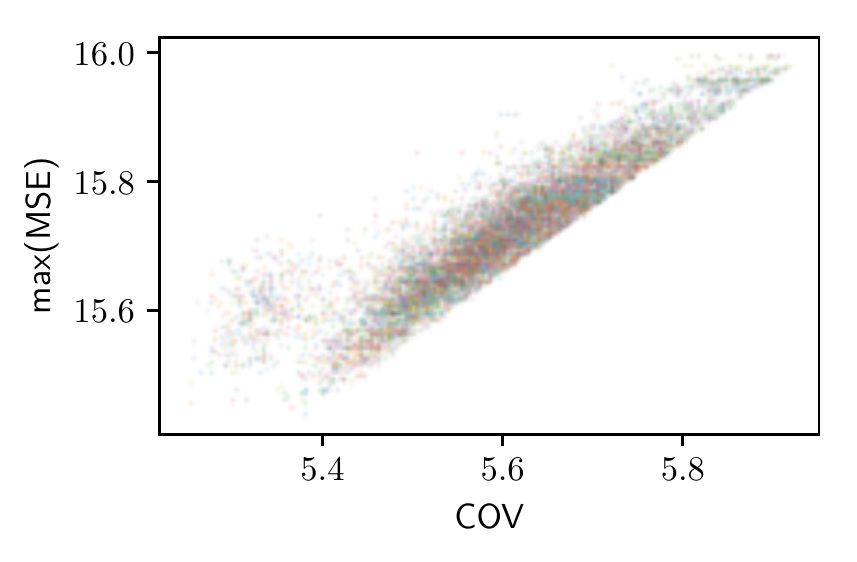}
  \caption{}
  \label{fig:scttr_cov_maxMSE_WS}
\end{subfigure}
\begin{subfigure}{.32\linewidth}
  \centering
\includegraphics[width=\linewidth]{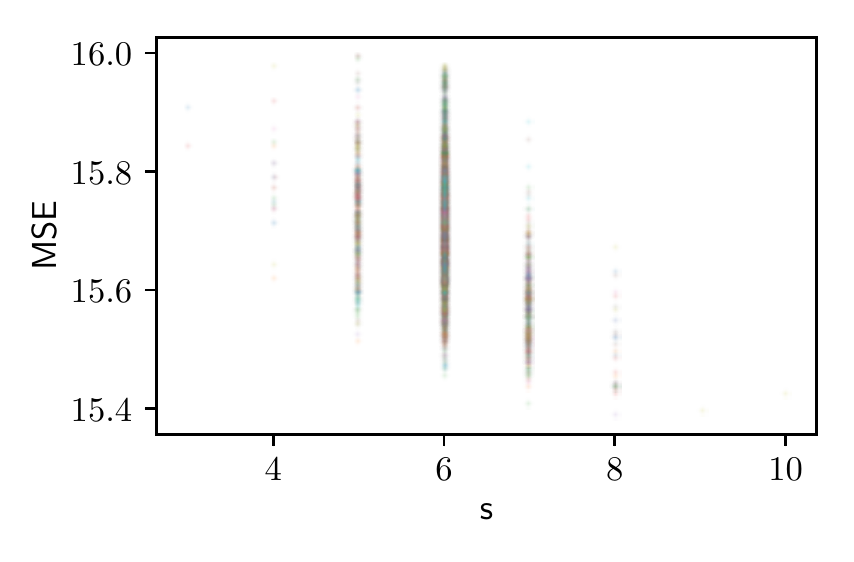}
  \caption{}
  \label{fig:scttr_MSE_deg_WS}
\end{subfigure}
\caption{ \textbf{(a)} Scatter plot of COV of a link vs. the minimum of MSE over the two ends of that link, \textbf{(b)} Scatter plot of COV of a link vs. the maximum of MSE over two ends of that link, \textbf{(c)} Scatter plot of MSE vs. degree of the nodes s. These figures are the result of simulations for 30 different realizations of SW networks with size 100, rewiring probability of 0.05 and mean degree 6.}
\label{fig:rndm_net_WS}
\end{figure}

Finally, we employ the Watts-Strogatz model \cite{watts1998collective} to generate 30 realizations of Small-World (SW) networks with size 100, rewiring probability 0.05 and mean degree 6. Fig. \ref{fig:rndm_net_WS} shows the results for this topology similar to the previous cases.

\begin{figure}
\centering
\includegraphics[width=0.4\linewidth]{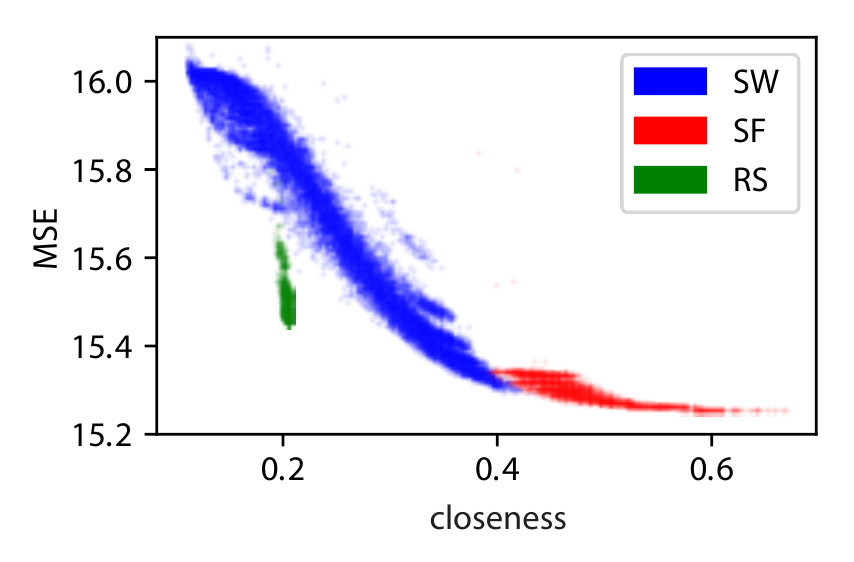}
\caption{Scatter plot of MSE of each cell vs. its closeness centrality for all realizations of three different topologies.
\label{fig:MSE_clsnss}}
\end{figure}

The results of simulations of the random networks suggest that independent of topology, a strong correlation exists between COV of errors of two neighboring cells and their MSE of estimation which indicates that when cells have larger MSE (i.e. worse estimation) they are more correlated to their neighbors. Moreover, the scatter plot of MSE of each cell vs. its closeness centrality (defined as the average of distance from the given node to all other nodes \cite{albert2002statistical}) for all realizations of the three different topologies shows an anti-correlation, as one can see in Fig. \ref{fig:MSE_clsnss}, indicating that higher closeness improves the quality of estimation.
 
\end{document}